\documentclass[a4paper,twocolumn,11pt,accepted=2026-02-23]{quantumarticle}
\pdfoutput=1
\usepackage[utf8]{inputenc}
\usepackage[english]{babel}
\usepackage[T1]{fontenc}
\usepackage{amsmath}
\usepackage{hyperref}
\usepackage{supertabular}

\usepackage{physics}
\usepackage{amssymb,amsmath,dsfont}
\usepackage{graphicx}
\usepackage[numbers,sort&compress]{natbib}
\usepackage{enumitem}
\usepackage{svg}
\usepackage{bm}
\usepackage{xcolor}
\usepackage{bm}
\usepackage{csvsimple}
\usepackage{booktabs}
\usepackage[caption=false]{subfig}
\usepackage{ulem}
\usepackage[capitalise]{cleveref}
\usepackage{comment}
\usepackage{braket}

\newcommand{\mean}[1]{\langle {#1} \rangle}

\newcommand{\balpha}{\ensuremath{{\bm \alpha}}}
\newcommand{\bmu}{\ensuremath{{\bm \mu}}}

\newcommand{\cor}[1]{\langle {#1} \rangle}

\begin{document}
\title{Automated generation of photonic circuits for Bell tests with homodyne measurements}

\author{Corentin Lanore}
\affiliation{Universit\'e Paris-Saclay, CEA, CNRS, Institut de physique th\'eorique, 91191, Gif-sur-Yvette, France}
\author{Federico Grasselli}
\affiliation{Universit\'e Paris-Saclay, CEA, CNRS, Institut de physique th\'eorique, 91191, Gif-sur-Yvette, France}
\affiliation{Leonardo Innovation Labs – Quantum Technologies, Via Tiburtina km 12,400, 00131 Rome, Italy}
\author{Xavier Valcarce}
\email{xavier.valcarce@ipht.fr}
\affiliation{Universit\'e Paris-Saclay, CEA, CNRS, Institut de physique th\'eorique, 91191, Gif-sur-Yvette, France}
\author{Jean-Daniel Bancal}
\affiliation{Universit\'e Paris-Saclay, CEA, CNRS, Institut de physique th\'eorique, 91191, Gif-sur-Yvette, France}
\author{Nicolas Sangouard}
\affiliation{Universit\'e Paris-Saclay, CEA, CNRS, Institut de physique th\'eorique, 91191, Gif-sur-Yvette, France}

\begin{abstract}
Nonlocal quantum realizations, certified by the violation of a Bell inequality, are core resources for device-independent quantum information processing. 
Although proof-of-principle experiments demonstrating device-independent quantum information processing have already been reported, identifying physical platforms that are realistically closer to practical, viable devices remains a significant challenge. 
In this work, we present an automated framework for designing photonic implementations of nonlocal realizations using homodyne detections and quantum state heralding.
Combining deep reinforcement learning and efficient simulations of quantum optical processes, our method generates photonic circuits that achieve significant violations of the Clauser-Horne-Shimony-Holt inequality.
In particular, we find an experimental setup, robust to losses, that yields a CHSH violation of \boldsymbol{$2.068$} with \boldsymbol{$3.9$}~dB and \boldsymbol{$0.008$}~dB squeezed light sources and two beam splitters. 
\end{abstract}

\maketitle

\section{Introduction} 
\label{sec:intro}

As demonstrated by the violation of Bell inequalities, quantum mechanics allows for entangled particles to exhibit nonlocal correlations that cannot be explained by any local hidden variable model~\cite{Bell1964,Brunner2014}.
Initially studied to probe the foundations of quantum theory, nonlocality has now found applications in the device-independent (DI) processing of quantum information~\cite{ArnonFriedman2020,Valcarce2023b}. 
Most notably, nonlocality enables the certification of quantum resources via self-testing~\cite{Mayers2004,Supic2020} and it is  necessary to derive device-independent security proofs for quantum key distribution~\cite{Acin2007, Pironio2009, Vazirani2014, Sekatski2021, Zapatero2019, Schwonnek2021, Zapatero2023} and to demonstrate the device-independent generation of quantum random numbers~\cite{Acin2016}. 

The nonlocality certified by a violation of the Clauser-Horne-Shimony-Holt (CHSH) inequality~\cite{CHSH1969}, a bipartite Bell inequality, stands at the core of numerous DI protocols. Loophole-free violations of the CHSH inequality have been accomplished using a variety of platforms, from NV-centers~\cite{Hensen2015} and photonic setup~\cite{Christensen13,Shalm15,Giustina15,Liu18,Shen18} to neutral atoms~\cite{Rosenfeld17} and superconducting circuits~\cite{Storz2023}. Efforts aiming to generate high CHSH violations at high rates have culminated with the first distribution of a device-independent key using trapped ions~\cite{Nadlinger2022}. Results have also been obtained to extend DIQKD over hundreds of meters with single atoms~\cite{Zhang2022}. On-going efforts~\cite{Liu2022} aim at implementing device-independent quantum information processing with a purely photonic platform – a platform where optical modes are entangled, manipulated and detected – which is plausibly closer to what is expected for a commercial device. 

With their high efficiency, low noise, high bandwidth and capability to operate at telecom wavelengths at room temperature, homodyne detectors are a natural candidate for implementing practical Bell tests.
Coupled with a fully photonic circuit for state generation and manipulation, Bell tests may be implemented using integrated on-chip devices.
Encouragingly, it has been shown that high or even maximal CHSH violations can be achieved with homodyne measurements~\cite{Gilchrist1999,Munro1999,Wenger2003,Oudot2024}.
However, these proposals require tailored states whose experimental realizations is either unknown or, in the best of cases, very challenging.
Specifically, the proposal of the physical process required to implement the state yielding a maximal CHSH score $\mathcal{B}$ proposed in \cite{Wenger2003} is very far fetched.
Likewise, the state producing a violation of $\mathcal{B}=2.14$ in \cite{Oudot2024} has a challenging implementation requiring six bosonic modes, each with high squeezing value, and four single-photon detectors.
Meanwhile, the best known CHSH score based on a realistic optical setup, which uses standard optics equipment and homodyne detectors, achieves a CHSH score of $\mathcal{B}\approx 2.048$~\cite{Garcia2004,Garcia2005}.
This score is, however, below the threshold required by some DI protocols. 
A notable example is self-testing the singlet state, which is infeasible for CHSH scores less than $2.051$~\cite{Valcarce2020}, while the most advanced protocol requires a score $\gtrapprox 2.106$~\cite{Kaniewski2016}.
Therefore, designing realistic optical setups with homodyne detection leading to higher CHSH scores remains an open problem.

With advances in integrated photonic circuits~\cite{Tanzilli2012,Pelucchi2021}, the implementation of circuits with a high number of elements and modes becomes possible, thus increasing the set of accessible optical states for CHSH tests.
It is however challenging to discover suitable optical circuits, as the number of possible circuits grows exponentially with the number of optical elements considered.

In this work, we tackle circuit discovery with an automated approach that combines a search algorithm with an efficient simulation of optical circuits based on the Gaussian representation of quantum states. 
Specifically, we consider the design of practical photonic CHSH tests, characterized by standard optical components and homodyne measurements.
Among the search strategies we employ, one is based on deep reinforcement learning, which has proven highly effective in identifying relevant elements in large and complex exploration spaces~\cite{Mnih2015,Mnih2016,Haarnoja2018,Schulman2015,Schulman2017} and has already been successfully applied to a variety of quantum physics tasks~\cite{Carleo2019,Biamonte2017,Dunjko2018,Krenn2020}, including the design of optical circuits~\cite{Krenn2016,Melnikov2020,Krenn2021,Valcarce2023}.
We also contrast performances of reinforcement learning with that of a simple random search over the same circuit space.

Our approach allows for the exploration of photonic circuits in different configurations by enforcing constraints on, e.g., the number of optical modes or the set of available optical elements.
This can narrow the search to specific sets of photonic circuits that reflect experimental constraints or leverage knowledge on photonic systems. 
We find numerous optical circuits yielding significant CHSH scores up to $\mathcal{B}\approx 2.076$.
In particular, we propose a simple setup composed of four bosonic modes and four optical components that produces a CHSH score of $\mathcal{B}\approx 2.068$. 
This setup yields CHSH violations for parties separated by more than $8$ kilometers, while requiring at most $3.9$~dB of squeezed light. 
We believe this proposed setup could lead to a first Bell test with homodyne measurements and pave the way for more practical implementations of DI protocols.

\section{Preliminaries}
\label{sec:preliminaries}

\subsection{The CHSH Inequality}

Bell tests allow detecting nonlocal quantum correlations by ruling out local hidden variable models for these correlations.
In this manuscript, we consider a Bell test involving two parties, Alice and Bob, and subdivided into rounds.
In each round, Alice picks a measurement setting by selecting one of two inputs $x\in\{0,1\}$, and collects the measurement outcome $a\in\{1,-1\}$.
Similarly, Bob chooses a setting $y\in\{0,1\}$ and obtains an outcome $b\in\{1,-1\}$.
With sufficiently many rounds, the parties can estimate the correlators
\begin{equation}
\cor{A_x B_y}= P(a=b|x,y)-P(a \neq b |x,y), \label{correlator}
\end{equation}
where $P(a=b|x,y)$ (resp. $P(a\neq b|x,y)$) is the probability that the outcomes $a$ and $b$ are equal (differ), given inputs $x$ for Alice and $y$ for Bob. 

In the case where the outcome correlations are compatible with a local hidden variable model, the Clauser-Horne-Shimony-Holt score
\begin{equation}
\mathcal{B} = \abs{\cor{A_0 B_0} + \cor{A_0 B_1} + \cor{A_1 B_0} - \cor{A_1 B_1}} \label{CHSH}
\end{equation}
satisfies the CHSH inequality~\cite{CHSH1969}
\begin{equation}
    \mathcal{B} \leq 2.
\end{equation}

Interestingly, quantum mechanics allows for violations of the CHSH inequality. This can be achieved using incompatible measurements on a shared entangled state~\cite{Brunner2014}. 
In particular, by performing appropriate projective measurements on two-qubit maximally entangled states, the CHSH score can reach the maximum quantum value of $\mathcal{B}=2\sqrt{2}$, known as the Tsirelson bound~\cite{Tsirelson1987}. 

\subsection{Photonic Circuits}

As we focus on practical photonic implementations of Bell tests, we first revisit the relevant concepts of quantum optics. 
After a brief introduction to photonics, we present Gaussian states and operations that can be accessed using standard optical components. 
We then review homodyne measurements and the binning of their outcomes, which are used to compute the CHSH score. 
Finally, we highlight the heralding of quantum states based on detection events with threshold detectors.
Further details are reported in \cref{app:quantopt}.

\medbreak

\paragraph{Photonics} 
The quadrature field operators of a bosonic mode are expressed in terms of ladder operators using $\hat{x} =(\hat{a}^\dagger + \hat{a})/2$ and $\hat{p} = i(\hat{a}^\dagger - \hat{a})/2$, such that $[\hat{x},\hat{p}]=i/2$ holds. 
Given $N$ bosonic modes, we arrange these operators in a vector $\mathbf{\hat{q}} = (\hat x_1, \hat p_1,\ldots, \hat x_n, \hat p_n)$.

\medbreak

\paragraph{Gaussian states}
\label{seq:GaussState}
Let $\hat{\rho}$ be an $N$-mode bosonic state. 
We define its displacement vector $\bmu$ to be composed of elements $\mu_i = \tr[\hat{q}_i \hat{\rho}]$, and its covariance matrix $\Sigma$ to have elements $\Sigma_{i,j} =\tr[(\hat{q_i} \hat{q_j} + \hat{q_j} \hat{q_i})\hat{\rho}]/2 - \mu_i \mu_j$.
Gaussian states are a class of quantum states which are completely characterized by their displacement vector and covariance matrix.

\medbreak

\paragraph{Gaussian operations}
\label{seq:GaussOp}
Unitary transformations that map Gaussian states into other Gaussian states are called Gaussian operations or Gaussian gates. 
These operations can be described directly in terms of their action on the displacement vector and the covariance matrix 
	\begin{align}
		\bmu &\mapsto M \bmu + \mathbf{d} \\
		\Sigma &\mapsto M \Sigma M^\dagger,
	\end{align}
where $M$ is a symplectic matrix. 
In this manuscript we consider four types of gates: two passive operations, beam splitters, $\hat{B}(\theta)$ and phase shifters $\hat{R}(\theta)$, as well as two active ones, namely single-mode and two-mode squeezers, labeled $\hat{S}_1(r)$ and $\hat{S}_2(r)$ respectively.
Note that these operations are all characterized by $\mathbf{d}=\vec{0}$, therefore leaving the displacement vector invariant if applied to a state with $\bmu=\vec{0}$.
The exact expressions of these transformations are reported in \cref{table:GaussianOperations} of \cref{sec:GaussianOperations}.

\medbreak

\paragraph{Homodyne measurements} 

Homodyne measurements are measurements of quadratures of the bosonic field.
In the Bell scenario we consider, Alice and Bob measure the quadratures of bosonic mode 1 and 2, respectively, with a randomly chosen measurement setting $x$ and $y$. Each setting corresponds to the measurement of a (fixed) 
rotated quadrature
\begin{align}
  \hat{x}^{\theta_x}_1 &= \cos(\theta_x)\hat{x}_1 + \sin(\theta_x)\hat{p}_1, \\
  \hat{x}^{\phi_y}_2 &= \cos(\phi_y)\hat{x}_2 + \sin(\phi_y)\hat{p}_2.
\end{align}
Such measurements can be implemented by a measurement of the $\hat{x}_i$ quadrature, preceded by a phase shifter with phase $\theta_x$ or $\phi_y$, accordingly.
The joint probability distribution over the continuous spectra of the two parties' measured quadratures is given by
\begin{equation}
    P(x^{\theta_x}_1 , x^{\phi_y}_2) = \iint_{-\infty}^{\infty} \text{d}p_1 \text{d}p_2 \tilde{W}^{xy}_{12} (x_1,p_1,x_2,p_2)
\end{equation}
where $\tilde{W}^{xy}_{12}$ is the Wigner function of the two-mode bosonic state shared by Alice and Bob, after local phase shifts with angle $\theta_x$ on Alice's mode and angle $\phi_y$ on Bob's mode.

As the CHSH scenario requires binary outcomes, we bin the continuous spectra using sign binning. This leads to the expression
\begin{equation}
    \cor{A_x B_y}= \iint_{-\infty}^{\infty} \mathrm{sign}(x^{\theta_x}_1 x^{\phi_y}_2) P(x^{\theta_x}_1,x^{\phi_y}_2) \mathrm{d}x^{\theta_x}_1 \mathrm{d}x^{\phi_y}_2
\end{equation}
for the correlators defined in \cref{correlator}. In \cref{app:chsh_computation}, we detail how we numerically compute the above expression.

\medbreak

\paragraph{Heralding quantum states}

Gaussian states and homodyne measurements alone are not suited for Bell inequality violations. 
Indeed, the statistics generated by Gaussian states and homodyne measurements always admit a local hidden variable model~\cite{Bell1986,Jabbour2023}, as the Wigner functions of both the state and the observables are non-negative. 
Conversely, non-Gaussian measurements such as photon-counting devices are described by Wigner functions that can be negative, thereby allowing for Bell violations with Gaussian states~\cite{Kuzmich2000}. In this work we take the opposite avenue to Bell violations, i.e.~we perform homodyne measurements on states with negative Wigner functions.

To obtain states with negative Wigner functions, we herald Gaussian states using threshold detectors. In particular, a heralded two-mode state $\rho_{12}$ can be obtained from an $N$-mode Gaussian state by measuring the presence of one (or more) photons in the other $N-2$ modes~\cite{Walschaers2021}. The resulting state can be written as a linear combination of Gaussian states
\begin{equation}
	\hat{\rho}_{12} = \sum_{k=1}^{2^{N-2}} \omega_k \hat{\rho}^k_{12}, 
	\label{pseudo-gaussian-states}
\end{equation}
with coefficients $\omega_k \in \mathbb{R}$ satisfying\footnote{Note that the coefficients $\omega_k$ need to satisfy additional constraints in order to ensure the positivity of the state $\hat{\rho}_{12}$.} $\sum_k \omega_k =1$, and where $\hat{\rho}^k_{12}$ are normalized two-mode Gaussian states. 
Since the Wigner function is linear in the density operator, the Wigner function of such heralded state may become negative in part of the phase space when some of the coefficients $\omega_k$ are negative.
Further details on the expression of heralded states and their creation with threshold detectors are given in \cref{sec:heralding}.

\subsection{Deep Reinforcement Learning}

Reinforcement learning (RL) is one of the three paradigms of machine learning. 
It is characterized by a trial-and-error approach to problem solving, which does not require pre-existing data for the training process.
Instead, RL relies on an agent interacting with an environment and receiving feedback in the form of rewards, which are based on the impact of the agent's actions. 
Through these interactions, the agent learns a policy, i.e.~a probabilistic strategy which aims at maximizing the cumulative reward over time.

The RL training process consists of a sequence of episodes, where each episode is composed of a series of steps.
At each step, the agent observes the current state of the environment, $s$, and selects an action, $a$, by sampling from the policy $\pi(a|s)$  -- the probability to take action $a$ given a state $s$.
The action is executed on the environment and a reward $r$ is returned to the agent.
When a given termination condition is met, e.g.~a maximum number of steps, the episode stops and the environment is reset to its initial state.
After a given number of episodes, the policy is optimized by learning from the past interactions with the environment (the sets of states, actions and respective rewards) with the aim of maximizing the cumulative rewards.

Policy optimization in RL can be tackled by a variety of algorithms. 
In this work we focus on the proximal policy optimization (PPO), introduced in~\cite{Schulman2017}, which balances exploration and exploitation while maintaining a stable and sample-efficient training.
In the PPO algorithm, the policy is optimized by iteratively updating it using a clipped surrogate objective function, which prevents drastic updates hence ensuring stability.
More specifically, PPO uses two neural networks: the policy network, which outputs a probability distribution $\pi(a|s)$ over actions $a$ conditioned on a state $s$, and the value network, which returns an estimation $V(s)$ of the expected cumulative reward from a state $s$.
Both networks take as input the current state of the environment $s$.
At each policy update, the weights of the policy network and the value network are updated to match the learned optimal policy and the received cumulative reward, respectively.

\section{Automated generation of photonic Bell tests} 
\label{sec:RL}

We employ RL, in combination with efficient circuit simulation and numerical optimization, to automate the generation of practical photonic circuits yielding CHSH violations.

We provide the full codebase implementing the environment, agent, and circuit-exploration strategies described in this section in an online repository~\cite{code_repo}.

\subsection{Photonic Bell tests as a learning task}

In the following, we formulate the tasks of finding photonic Bell tests in the \textit{agent-environment} framework.
More specifically, we present the environment implementations and the set of actions the agent can interact with, the state and reward received as feedback, as well as an episode routine from initial conditions to termination.
Note that this framework is not restricted to reinforcement learning agents and can be used by other search algorithms.

The environment is chosen to be an $N$-mode optical circuit, where the first mode is sent to Alice and the second to Bob, while the remaining $N-2$ modes are heralded during state preparation.
As an initial condition, each mode is set to the vacuum state.
The agent acts on the environment by appending photonic gates to the optical circuit. 
Each action is identified by the selected gate, either a beam splitter $\hat{B}(\theta)$, phase shifter $\hat{R}(\theta)$, single-mode squeezer $\hat{S}_1(r)$ or two-mode squeezer $\hat{S}_2(r)$, and the mode or modes on which the gate is applied. 
At this stage, the parameters of each photonic gate are fixed to a predefined value in order to limit the size of the action space during training.
After each action on the environment, the agent receives in return a state and a reward. 
Finally, once the circuit depth (i.e.~number of gates) reaches a maximum predefined value $n_{\rm circuit}$, we terminate the episode and reset the photonic circuit to empty modes.

The state returned by the environment is given by the photonic state shared between Alice and Bob.
This state is computed by applying the current gates in the circuit and by heralding the state of the first two modes on a photon detection in each of the last $N-2$ modes. 
This computation is efficiently carried out numerically by the code which is made available in Ref.~\cite{Valcarce2021}.
As the heralded state can be seen as a linear combination of $2^{N-2}$ Gaussian states, as in \cref{pseudo-gaussian-states}, we represent it as a vector of $2^{N-2}(10+1)$ real elements.
More specifically, the vector is a concatenation over all $k$ of the $10$ elements of the covariance matrix of each Gaussian state, c.f. \cref{sec:gaussianstates}, $\hat\rho_{12}^k$, followed by the coefficients $\omega_k$ of the linear combination (remember that the Gaussian parameter $\mu$ is always zero here so we do not need to store it).
Note that if the photon detection probability in one of the heralded modes is below a certain threshold, set to $10^{-10}$, the heralding is considered failed to avoid numerical errors (note that this still allows for a global heralding probability smaller than $10^{-10}$ in the presence of several heralding modes). 
In this case, the state passed to the agent is a trivial two-mode vacuum state. 

As the cumulative reward over the course of an episode is the quantity the policy aims at maximizing, we give a reward of zero for all but the last step.
For the last step, i.e.~once the circuit is composed of $n_\text{circuit}$ gates, we compute the CHSH score of the total circuit by maximizing $\mathcal{B}$ over all the optical gates' parameters with the measurement setting $\{\theta_0,\theta_1\}=\{0,\pi/2\}$ and $\{\phi_0,\phi_1\}=\{-\pi/4,\pi/4\}$, and using sign binning. This is done using the Nelder-Mead algorithm. 
The reward received by the agent is an increasing function of the CHSH score
\begin{align}
    r(\mathcal{B}) = \left\lbrace \begin{array}{ll}
        \frac{\mathcal{B}}{4} -1 & \mbox{if } \mathcal{B} < 2 \\[1ex]
        \exp\left[ 10 \ln(2) (\mathcal{B}-2)\right] -1  & \mbox{otherwise}.
    \end{array} \right.
\end{align}
This reward function was designed empirically to bias the agent toward circuits that violates the CHSH inequality. 
In particular, non-violating circuits, i.e. $\mathcal{B}<2$, receive only a weak, linear reward, whereas circuits achieving a violation are assigned a reward that grows exponentially with the CHSH score. 
This exponential scaling encourages the agent not only to discover violating circuits, but also to favor improvements in the score, even when such improvements are marginal.

\subsection{Learning to design Bell tests}

A PPO agent is entrusted with learning to design practical photonic Bell tests.
This makes use of a two-headed four-layer neural network, for the policy and value network.
The input layer encodes the parameters of the heralded state while the neurons of the output layer represent different actions for the policy head, and the single neuron of the value-head returns the scalar $V(s)$. 
The number of neurons in the two hidden layers is chosen depending on the size of the state parametrization and the size of the action space. 
For example, we choose $150$ and $90$ neurons in the first and second hidden layer, respectively, when testing circuits with $N=6$ modes where we restrict the agent's actions to only passive gates. We instead choose 45 and 30 neurons for $N=4$ modes and arbitrary gates.
We use the PPO agent implementation of the Julia Reinforcement Learning package~\cite{Tian2020}, and rely on the default hyperparameters, found in \cref{tab:hyperparameters}, except for the update frequency and the trajectory capacity, which are set to 64 and 256, respectively.

\subsection{Exploration and learning strategies}\label{sec:strategies}

In order to maximize the efficiency of the learning process, we refine our circuit construction model within the framework defined above by identifying several learning strategies. These strategies are motivated by physical and machine learning insights as well as initial benchmarks, and amount to choices of circuit initialization and restrictions on the sets of gates available to the RL agent. An initial study of these strategies by random search allows us to launch the RL algorithm only on the most suitable strategies. Additionally, we consider several heralding schemes as detailed below.

The first circuit-building strategy we consider is the most general one in which the circuit is initialized in the vacuum state and the agent is allowed to pick any of the four gates ($\hat{B}(\theta)$, $\hat{R}(\theta)$, $\hat{S}_1(r)$ and $\hat{S}_2(r)$) in every interaction with the environment. This strategy has the merit of letting the agent free to choose the optimal circuit configuration, but it comes at the cost of a large action space. 
Moreover, this strategy permits circuit configurations where multiple squeezers act on the same mode yielding large squeezing parameters. 
Actually, the agent is probably incentivized to abuse of squeezers as they can increase the amount of entanglement and enhance the CHSH violations. 
However, the use of sequential squeezing may be hard to achieve experimentally, and it can easily cause numerical errors that lead to unrealistic CHSH scores.
Another shortcoming of this strategy is that the agent will inevitably attempt to place passive gates at the beginning of the circuit during training, which would have no effect since they act on the vacuum, thereby decreasing the circuit depth available to the agent. For these reasons, we define additional strategies which avoid these limitations.

Concretely, we consider four supplementary strategies in which the circuit is initialized with some fixed active gates that generate photons (single- or two-mode squeezers) and the agent is only allowed to choose gates that are passive (phase shifter and beam splitter) for the rest of the episode. The specific initialization of each strategy is described in \cref{app:circuits}. These new strategies avoid the unnecessary action choices present in the first strategy because the circuit is initialized in a state containing photons in each mode. 
Additionally, the risk of too large squeezing parameters is reduced by the presence of at most one squeezer per mode.
This constraint also favors experimentally realistic setups as chaining squeezing operations appears challenging.
It should be noted that despite limiting the agent to passive gates only, these strategies are not restrictive.
Indeed, the Euler decomposition (c.f. \cref{app:quantopt}) guarantees that an arbitrary circuit initialized in the vacuum can always be reproduced by a circuit initialized with squeezers and completed by only passive gates. 
Finally, by reducing the size of the action space, such strategies can favour the learning process of the agent.

In order to identify the most suitable of these exploration strategies, we test them on random circuits. Namely, we construct circuits by appending successively uniformly chosen random gates from the chosen set until the circuit depth $n_\mathrm{circuit}$ is reached. With no reliance on neural networks, such a random search is light on computational resource and, therefore, allows for the automated exploration of many more circuits in a given time. In particular, a random exploration may provide a circuit yielding a suitable CHSH score. 
This exploration is however unstructured and may become less effective as the size of the search space grows.
In our case, we use the result of this investigation to identify the most suitable exploration strategies for reinforcement learning (see Appendix B for more details).

Further preliminary tests also show that circuits with $N=3$ and $N=5$ modes return weak results. We thus focus our automated search on circuits with $N=4$ and $N=6$ modes.

Finally, for each considered circuit building strategy is explored using each of two heralding mechanisms for the last $N-2$ modes.
In the first scheme, we select states based on detection events from threshold detectors applied to each of the last $N-2$ modes. 
A threshold detector differentiates between vacuum and the presence of one or more photons. 
In the second heralding scheme, we aim to better approximate the single-photon subtraction operation, which is known to introduce non-Gaussianity that is essential for generating nonlocality~\cite{Walschaers2021}. 
This is achieved using an unbalanced beam splitter with high transmittivity, along with a threshold detector at each output port. 
We select events where a click occurs in the reflected port and a vacuum is detected in the transmitted port. 
For further details on both heralding schemes, we refer the reader to \cref{app:circuits}.

\section{Results}
\label{sec:results}

\begin{figure}[t]
    \centering
    \includegraphics[width=8cm]{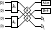}
    \caption{Optical circuit resulting from our automated search. The circuit is composed of two two-mode squeezers ($\hat{S}_2$) acting on modes $(1,2)$ and $(3,4)$ with respective parameters $r = 0.00096$ and $r = 0.44993$, followed by two beam splitters ($\hat{B}$) on modes $(1,3)$ and $(2,4)$, with respective parameters $\theta = 1.50272$ and $\theta = 1.63856$. The state of modes $(1,2)$, shared by Alice and Bob, is heralded on a click in both the threshold detectors of modes $(3,4)$. Alice and Bob measure their modes with homodyne measurements and can obtain an average CHSH score of $\mathcal{B}= 2.068$.}
    \label{fig:result1}
\end{figure}

In this section we discuss the optical circuits for CHSH violation obtained with our automated approach.
We then focus on a specific circuit, depicted in \cref{fig:result1}, which has the merit of being relatively simple while providing a significant CHSH score.
We analyse the resilience of this circuit to photon loss with distance between the parties, as well as its compatibility with inefficient detectors. 
To ensure reproducibility, the Julia notebook used to produce the analysis and plots of this section can be found online~\cite{code_repo}.

\medbreak

Using the selected strategies, the PPO agent produced several optical circuits admitting a significant violation of the CHSH inequality. In particular, when acting on $N=4$ modes, the agent found a circuit producing the score $\mathcal{B}=2.069$, a slightly larger value than the best score $\mathcal{B}=2.068$ obtained when using random search in this setting.

However, we also observe that in a given allocated time, the agent could not consistently outperform the random search. In presence of $N=6$ modes, for instance, the best circuit identified by the PPO approach reached $\mathcal{B}=2.072$, whereas random search achieved $\mathcal{B}=2.076$. It is worth noting that the corresponding circuit is significantly more convoluted, though, involving 42 gates for the random search compared to just 8 gates for the solution provided by the PPO agent.
This non-optimal CHSH score is partially due to the fact that our agent is trained with modest computing power and does not rely on a large neural network. 
Therefore, the agent cannot learn complex circuit patterns required to reach higher CHSH scores. 
Moreover, learning a suboptimal policy has the additional negative effect of reducing the exploration in later episodes, which further reduces the chances to find highly non-trivial circuits leading to higher CHSH scores. A summary of the circuits found by the PPO agent and by random search for the different strategies is provided in \cref{app:circuits}.

\medbreak

We now focus on the circuit depicted in \cref{fig:result1}.
This circuit appears as a good candidate for implementing a photonic Bell test with homodyne measurements.
It has been found by the PPO agent interacting with a circuit of $N=4$ modes initialized with two two-mode squeezers and is composed of only two additional beam splitters.
The third and fourth modes are used to herald the state using threshold detectors. 
This photonic circuit yields a CHSH score of $\mathcal{B}=2.068$ with optimal squeezing parameters corresponding to $3.9$~dB and $8 \cdot 10^{-3}$~dB of squeezing, respectively. 
The heralding probability of the state is $3\cdot 10^{-6}$.
Considering squeezers generating pulses at the MHz rate, the proposed circuit can prepare about $300$ states per second.
Therefore, the statistics for a Bell test can be gathered in a realistic amount of time, e.g.~$10^6$ shots in an hour.

\begin{figure}[t]
    \centering
    \includegraphics[width=8cm]{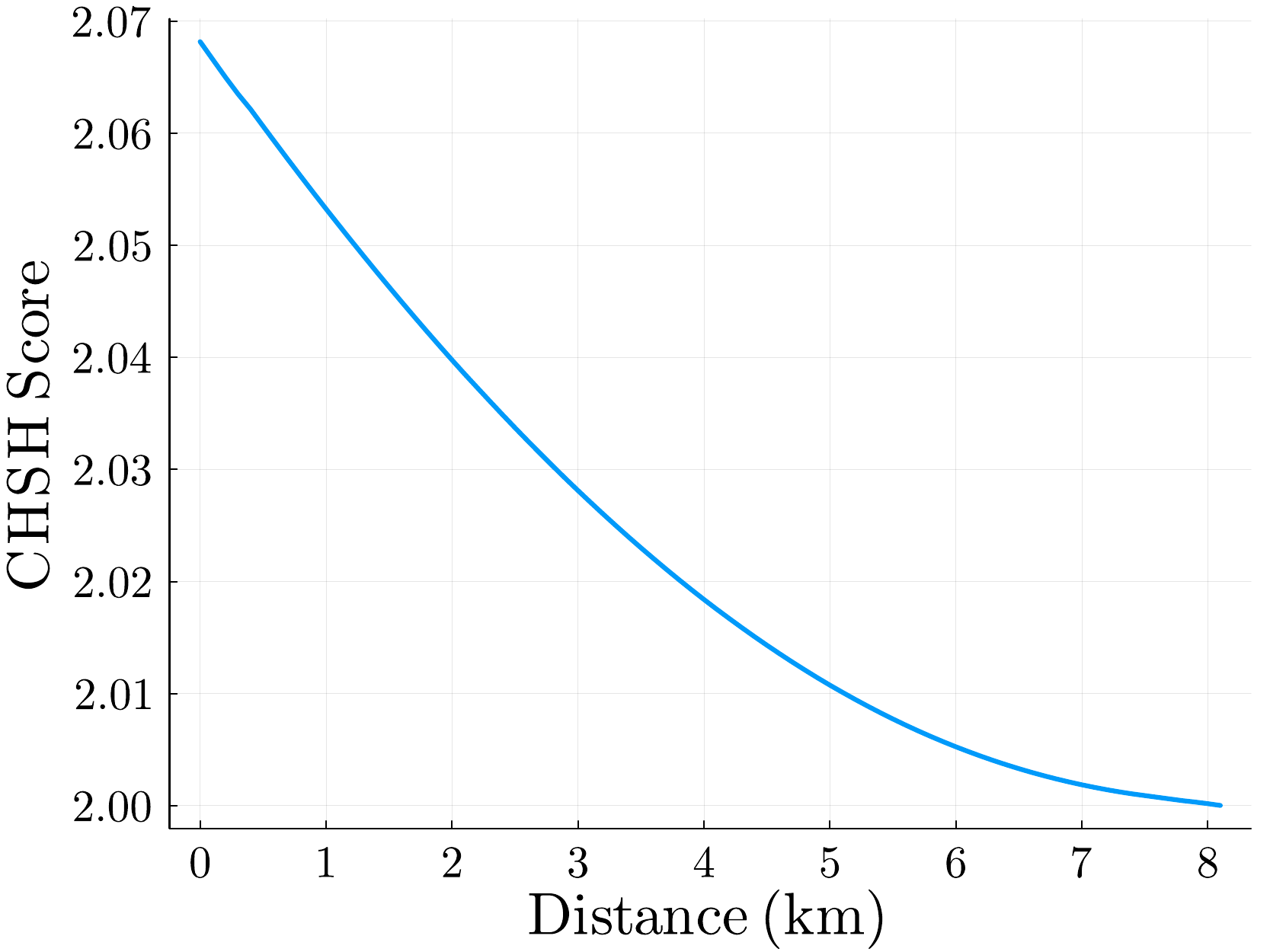}
    \caption{CHSH score with respect to the distance between Alice, where state preparation occurs, and Bob. We consider an optical fiber with a loss of $0.2$~dB/km, and numerically optimized the circuits' parameters for each step of $0.1$~km. As higher squeezing is beneficial for greater distance, we ensure squeezing of at most $10$dB to respect practical experimental range. }
    \label{fig:result2}
\end{figure}

\begin{figure}[ht]
    \centering
    \includegraphics[width=8cm]{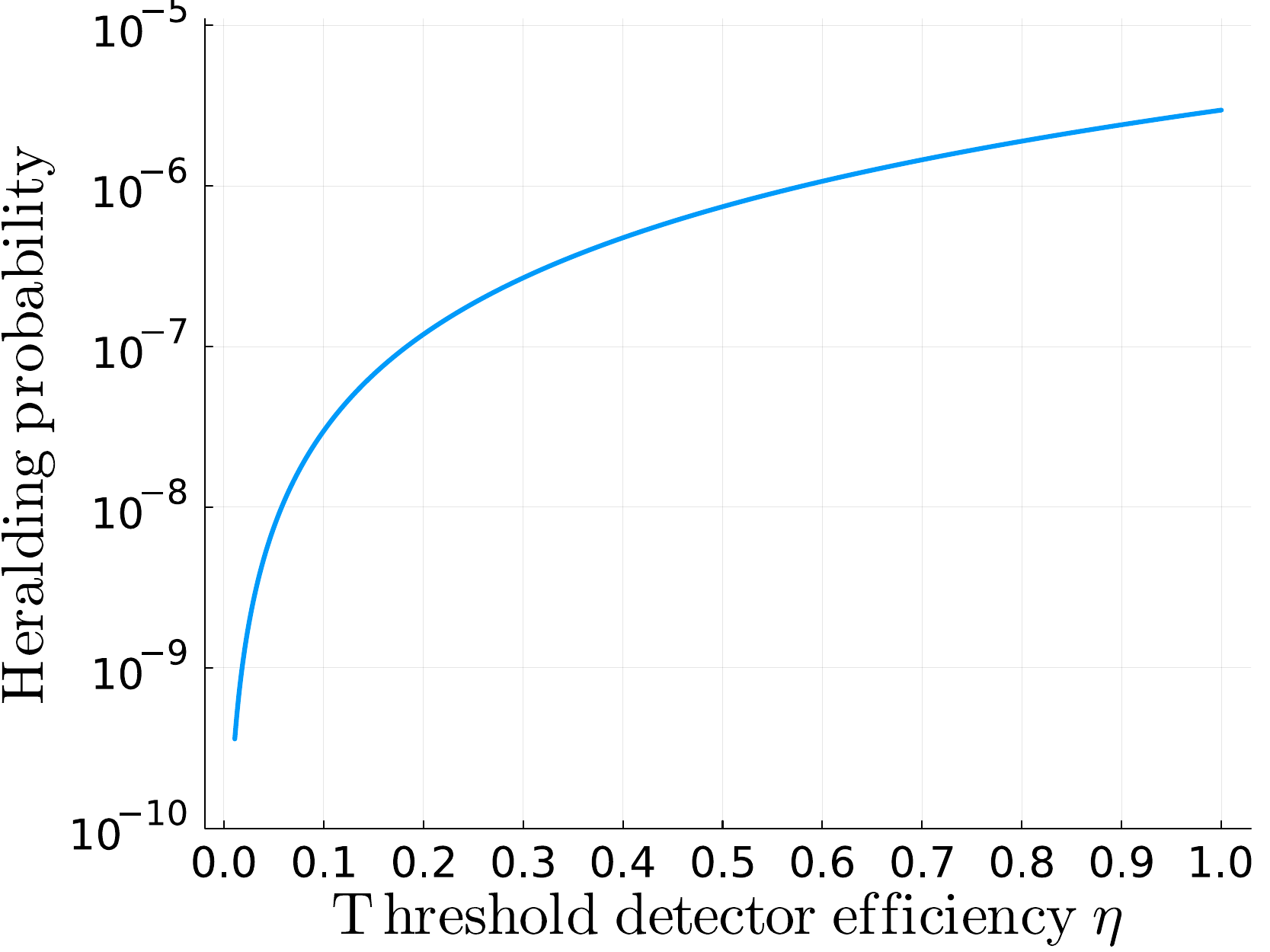}
    \caption{Evolution of the probability of a successful state heralding with respect to the efficiency of threshold detectors. }
    \label{fig:result4}
\end{figure}

\medbreak

We further study the impact of photon loss and inefficient detectors on the two most important metrics of the circuit, namely the CHSH score and the heralding probability.

To study the effect of photon loss, we assume that the optical circuit preparing and heralding the quantum state is located in Alice's laboratory.
Bob receives his part of the state, i.e the second mode, through an optical fiber link characterized by $0.2$~dB/km of attenuation. 
We assume that the optical signals are prepared at telecom wavelength, and, hence, that no conversions are necessary.
In \cref{fig:result2}, we plot the CHSH score as a function of the distance between Alice and Bob.
For every value of distance, we re-optimize the parameters of the gates to maximize the CHSH score. In \cref{tab:parameters} we provide the optimized value of these parameters with distance. We observe that a CHSH violation is still possible for distances up to $8.1$ km.
As for the heralding probability of the state, this is not affected by photon loss occurring in Alice's or Bob's modes.
See \cref{app:photonloss} for more details on this analysis.

We then analyse the effect of imperfect detectors on circuit performance.
Specifically, we assume that the threshold detectors used in the heralded modes have efficiency $\eta<1$. 
In \cref{fig:result4} we plot the heralding probability as a function of $\eta$, for fixed gate parameters optimal in the $\eta=1$ case (see \cref{app:ineffthreshold} for more details). 
We observe that the heralding probability decreases by at most one order of magnitude for efficiencies $\eta\geq 0.25$. 
Moreover, we observe that the efficiency of the detectors has barely any impact on the CHSH score.
Specifically, from a value $\mathcal{B} = 2.068$ at $\eta = 1.0$, the CHSH score approaches $\mathcal{B} = 2.067$ as $\eta \rightarrow 0$. 
This can be explained by the fact that the average number of photons reaching the threshold detectors is well below $1$. 
Indeed, if an inefficient threshold detector measures one mode out of a multi-photon entangled state, it detects each $n$-photon component of the state with a different probability, $1-(1-\eta)^n$, thereby modifying the photon number distribution of the unmeasured modes accordingly. 
This causes the conditional state of the unmeasured modes to depend, in principle, on the detector's efficiency. However, if the multi-photon components in the measured mode are suppressed, the dependence of the conditional state on the detector efficiency becomes negligible, and the only real consequence is a reduction in the heralding probability.

Given the negligible impact of detector efficiency on the CHSH score and its minor influence on the heralding probability, we conclude that highly efficient detectors, such as single-photon detectors operating at cryogenic temperatures, are not required for a successful Bell tests using the proposed photonic circuit. 
Combined with its robustness to losses, this approach may enable loophole-free Bell tests using homodyne measurement and basic threshold detectors.

\section{Conclusion} 
\label{sec:conclusion}

In this manuscript, we explore implementations for a photonic Bell inequality violation with homodyne measurements. 
We investigate practical optical setups, using standard optical components, with the aim of maximizing the violation of the CHSH inequality. 
This is achieved in an automated way thanks to reinforcement learning and random search, coupled with an efficient simulation framework of optical processes. 
By enforcing different sets of constraints on the explored photonic circuits, our approach leads to several circuits achieving a CHSH score higher than known proposals relying on homodyne measurements.
In particular, a fairly simple circuit of four optical modes, two of which are heralded with threshold detectors, and four optical components reaches a CHSH score of $2.068$. 
Through noise analysis, we have shown the resilience of the circuit against photon loss and its robustness to inefficient threshold detectors.
The proposed circuit may open a way towards the first loophole-free Bell test with homodyne measurements. 
More broadly, the method described in this manuscript may be applied to discover photonic circuits using homodyne measurements, that could implement quantum information protocols relying on nonlocality, such as device-independent quantum key distribution.

\begin{acknowledgments}
The authors acknowledge funding by the European Union’s Horizon Europe research and innovation program under the project “Quantum Security Networks Partnership” (QSNP, Grant Agreement No. 101114043) and by a French national quantum initiative managed by Agence Nationale de la Recherche in the framework of France 2030 with the reference ANR-22-PETQ-0009. FG did not contribute to this work on behalf of Leonardo S.p.A.
\end{acknowledgments}

\bibliography{references}

\begin{thebibliography}{10}

\bibitem{Bell1964}
J.~S. Bell.
\newblock ``On the {E}instein {P}odolsky {R}osen paradox''.
\newblock \href{https://dx.doi.org/10.1103/PhysicsPhysiqueFizika.1.195}{Physics
  Physique Fizika {\bf 1}, 195--200}~(1964).

\bibitem{Brunner2014}
Nicolas Brunner, Daniel Cavalcanti, Stefano Pironio, Valerio Scarani, and
  Stephanie Wehner.
\newblock ``Bell nonlocality''.
\newblock \href{https://dx.doi.org/10.1103/RevModPhys.86.419}{Rev. Mod. Phys.
  {\bf 86}, 419--478}~(2014).

\bibitem{ArnonFriedman2020}
Rotem Arnon-Friedman.
\newblock ``Device-independent quantum information processing: A simplified
  analysis''.
\newblock \href{https://dx.doi.org/10.1007/978-3-030-60231-4}{Springer
  International Publishing}. ~(2020).

\bibitem{Valcarce2023b}
Xavier Valcarce.
\newblock ``{Device-independent certification : quantum resources and quantum
  key distribution}''.
\newblock Theses.
\newblock {Universit{\'e} Paris-Saclay}.
\newblock ~(2023).
\newblock  url:~\url{https://theses.hal.science/tel-04132704}.

\bibitem{Mayers2004}
Dominic Mayers and Andrew Yao.
\newblock ``Self testing quantum apparatus''.
\newblock \href{https://dx.doi.org/10.48550/arXiv.quant-ph/0307205}{Quantum
  Info. Comput. {\bf 4}, 273–286}~(2004).

\bibitem{Supic2020}
Ivan {\v{S}}upi{\'{c}} and Joseph Bowles.
\newblock ``Self-testing of quantum systems: a review''.
\newblock \href{https://dx.doi.org/10.22331/q-2020-09-30-337}{{Quantum} {\bf
  4}, 337}~(2020).

\bibitem{Acin2007}
Antonio Ac\'{\i}n, Nicolas Brunner, Nicolas Gisin, Serge Massar, Stefano
  Pironio, and Valerio Scarani.
\newblock ``Device-independent security of quantum cryptography against
  collective attacks''.
\newblock \href{https://dx.doi.org/10.1103/PhysRevLett.98.230501}{Phys. Rev.
  Lett. {\bf 98}, 230501}~(2007).

\bibitem{Pironio2009}
Stefano Pironio, Antonio Acín, Nicolas Brunner, Nicolas Gisin, Serge Massar,
  and Valerio Scarani.
\newblock ``Device-independent quantum key distribution secure against
  collective attacks''.
\newblock \href{https://dx.doi.org/10.1088/1367-2630/11/4/045021}{New Journal
  of Physics {\bf 11}, 045021}~(2009).

\bibitem{Vazirani2014}
Umesh Vazirani and Thomas Vidick.
\newblock ``Fully device-independent quantum key distribution''.
\newblock \href{https://dx.doi.org/10.1103/PhysRevLett.113.140501}{Phys. Rev.
  Lett. {\bf 113}, 140501}~(2014).

\bibitem{Sekatski2021}
Pavel Sekatski, Jean-Daniel Bancal, Xavier Valcarce, Ernest Y.-Z. Tan, Renato
  Renner, and Nicolas Sangouard.
\newblock ``Device-independent quantum key distribution from generalized {CHSH}
  inequalities''.
\newblock \href{https://dx.doi.org/10.22331/q-2021-04-26-444}{Quantum {\bf 5},
  444}~(2021).

\bibitem{Zapatero2019}
Víctor Zapatero and Marcos Curty.
\newblock ``Long-distance device-independent quantum key distribution''.
\newblock \href{https://dx.doi.org/10.1038/s41598-019-53803-0}{Scientific
  Reports {\bf 9}, 17749}~(2019).

\bibitem{Schwonnek2021}
René Schwonnek, Koon~Tong Goh, Ignatius~W. Primaatmaja, Ernest Y.-Z. Tan,
  Ramona Wolf, Valerio Scarani, and Charles C.-W. Lim.
\newblock ``Device-independent quantum key distribution with random key
  basis''.
\newblock \href{https://dx.doi.org/10.1038/s41467-021-23147-3}{Nature
  Communications {\bf 12}, 2880}~(2021).

\bibitem{Zapatero2023}
Víctor Zapatero, Tim van Leent, Rotem Arnon-Friedman, Wen-Zhao Liu, Qiang
  Zhang, Harald Weinfurter, and Marcos Curty.
\newblock ``Advances in device-independent quantum key distribution''.
\newblock \href{https://dx.doi.org/10.1038/s41534-023-00684-x}{npj Quantum
  Information {\bf 9}, 10}~(2023).

\bibitem{Acin2016}
Antonio Acín and Lluis Masanes.
\newblock ``Certified randomness in quantum physics''.
\newblock \href{https://dx.doi.org/10.1038/nature20119}{Nature {\bf 540},
  213–219}~(2016).

\bibitem{CHSH1969}
John~F. Clauser, Michael~A. Horne, Abner Shimony, and Richard~A. Holt.
\newblock ``Proposed experiment to test local hidden-variable theories''.
\newblock \href{https://dx.doi.org/10.1103/PhysRevLett.23.880}{Phys. Rev. Lett.
  {\bf 23}, 880--884}~(1969).

\bibitem{Hensen2015}
B.~Hensen, H.~Bernien, A.~E. Dréau, A.~Reiserer, N.~Kalb, M.~S. Blok,
  J.~Ruitenberg, R.~F.~L. Vermeulen, R.~N. Schouten, C.~Abellán, W.~Amaya,
  V.~Pruneri, M.~W. Mitchell, M.~Markham, D.~J. Twitchen, D.~Elkouss,
  S.~Wehner, T.~H. Taminiau, and R.~Hanson.
\newblock ``Loophole-free {B}ell inequality violation using electron spins
  separated by 1.3 kilometres''.
\newblock \href{https://dx.doi.org/10.1038/nature15759}{Nature {\bf 526},
  682–686}~(2015).

\bibitem{Christensen13}
B.~G. Christensen, K.~T. McCusker, J.~B. Altepeter, B.~Calkins, T.~Gerrits,
  A.~E. Lita, A.~Miller, L.~K. Shalm, Y.~Zhang, S.~W. Nam, N.~Brunner, C.~C.~W.
  Lim, N.~Gisin, and P.~G. Kwiat.
\newblock ``Detection-loophole-free test of quantum nonlocality, and
  applications''.
\newblock \href{https://dx.doi.org/10.1103/PhysRevLett.111.130406}{Phys. Rev.
  Lett. {\bf 111}, 130406}~(2013).

\bibitem{Shalm15}
Lynden~K. Shalm, Evan Meyer-Scott, Bradley~G. Christensen, Peter Bierhorst,
  Michael~A. Wayne, Martin~J. Stevens, Thomas Gerrits, Scott Glancy, Deny~R.
  Hamel, Michael~S. Allman, Kevin~J. Coakley, Shellee~D. Dyer, Carson Hodge,
  Adriana~E. Lita, Varun~B. Verma, Camilla Lambrocco, Edward Tortorici, Alan~L.
  Migdall, Yanbao Zhang, Daniel~R. Kumor, William~H. Farr, Francesco Marsili,
  Matthew~D. Shaw, Jeffrey~A. Stern, Carlos Abell\'an, Waldimar Amaya, Valerio
  Pruneri, Thomas Jennewein, Morgan~W. Mitchell, Paul~G. Kwiat, Joshua~C.
  Bienfang, Richard~P. Mirin, Emanuel Knill, and Sae~Woo Nam.
\newblock ``Strong loophole-free test of local realism''.
\newblock \href{https://dx.doi.org/10.1103/PhysRevLett.115.250402}{Phys. Rev.
  Lett. {\bf 115}, 250402}~(2015).

\bibitem{Giustina15}
Marissa Giustina, Marijn A.~M. Versteegh, S\"oren Wengerowsky, Johannes
  Handsteiner, Armin Hochrainer, Kevin Phelan, Fabian Steinlechner, Johannes
  Kofler, Jan-\AA{}ke Larsson, Carlos Abell\'an, Waldimar Amaya, Valerio
  Pruneri, Morgan~W. Mitchell, J\"orn Beyer, Thomas Gerrits, Adriana~E. Lita,
  Lynden~K. Shalm, Sae~Woo Nam, Thomas Scheidl, Rupert Ursin, Bernhard
  Wittmann, and Anton Zeilinger.
\newblock ``Significant-loophole-free test of {B}ell's theorem with entangled
  photons''.
\newblock \href{https://dx.doi.org/10.1103/PhysRevLett.115.250401}{Phys. Rev.
  Lett. {\bf 115}, 250401}~(2015).

\bibitem{Liu18}
Yang Liu, Qi~Zhao, Ming-Han Li, Jian-Yu Guan, Yanbao Zhang, Bing Bai, Weijun
  Zhang, Wen-Zhao Liu, Cheng Wu, Xiao Yuan, Hao Li, W.~J. Munro, Zhen Wang,
  Lixing You, Jun Zhang, Xiongfeng Ma, Jingyun Fan, Qiang Zhang, and Jian-Wei
  Pan.
\newblock ``Device-independent quantum random-number generation''.
\newblock \href{https://dx.doi.org/10.1038/s41586-018-0559-3}{Nature {\bf 562},
  548--551}~(2018).

\bibitem{Shen18}
Lijiong Shen, Jianwei Lee, Le~Phuc Thinh, Jean-Daniel Bancal, Alessandro
  Cer\`e, Antia Lamas-Linares, Adriana Lita, Thomas Gerrits, Sae~Woo Nam,
  Valerio Scarani, and Christian Kurtsiefer.
\newblock ``Randomness extraction from {B}ell violation with continuous
  parametric down-conversion''.
\newblock \href{https://dx.doi.org/10.1103/PhysRevLett.121.150402}{Phys. Rev.
  Lett. {\bf 121}, 150402}~(2018).

\bibitem{Rosenfeld17}
Wenjamin Rosenfeld, Daniel Burchardt, Robert Garthoff, Kai Redeker, Norbert
  Ortegel, Markus Rau, and Harald Weinfurter.
\newblock ``Event-ready {B}ell test using entangled atoms simultaneously
  closing detection and locality loopholes''.
\newblock \href{https://dx.doi.org/10.1103/PhysRevLett.119.010402}{Phys. Rev.
  Lett. {\bf 119}, 010402}~(2017).

\bibitem{Storz2023}
Simon Storz, Josua Sch\"{a}r, Anatoly Kulikov, Paul Magnard, Philipp Kurpiers,
  Janis L\"{u}tolf, Theo Walter, Adrian Copetudo, Kevin Reuer, Abdulkadir Akin,
  Jean-Claude Besse, Mihai Gabureac, Graham~J. Norris, Andrés Rosario, Ferran
  Martin, José Martinez, Waldimar Amaya, Morgan~W. Mitchell, Carlos Abellan,
  Jean-Daniel Bancal, Nicolas Sangouard, Baptiste Royer, Alexandre Blais, and
  Andreas Wallraff.
\newblock ``Loophole-free {B}ell inequality violation with superconducting
  circuits''.
\newblock \href{https://dx.doi.org/10.1038/s41586-023-05885-0}{Nature {\bf
  617}, 265–270}~(2023).

\bibitem{Nadlinger2022}
D.~P. Nadlinger, P.~Drmota, B.~C. Nichol, G.~Araneda, D.~Main, R.~Srinivas,
  D.~M. Lucas, C.~J. Ballance, K.~Ivanov, E.~Y.-Z. Tan, P.~Sekatski, R.~L.
  Urbanke, R.~Renner, N.~Sangouard, and J.-D. Bancal.
\newblock ``Experimental quantum key distribution certified by
  {B}ell{\textquotesingle}s theorem''.
\newblock \href{https://dx.doi.org/10.1038/s41586-022-04941-5}{Nature {\bf
  607}, 682--686}~(2022).

\bibitem{Zhang2022}
Wei Zhang, Tim van Leent, Kai Redeker, Robert Garthoff, Ren{\'{e}} Schwonnek,
  Florian Fertig, Sebastian Eppelt, Wenjamin Rosenfeld, Valerio Scarani,
  Charles C.-W. Lim, and Harald Weinfurter.
\newblock ``A device-independent quantum key distribution system for distant
  users''.
\newblock \href{https://dx.doi.org/10.1038/s41586-022-04891-y}{Nature {\bf
  607}, 687--691}~(2022).

\bibitem{Liu2022}
Wen-Zhao Liu, Yu-Zhe Zhang, Yi-Zheng Zhen, Ming-Han Li, Yang Liu, Jingyun Fan,
  Feihu Xu, Qiang Zhang, and Jian-Wei Pan.
\newblock ``Toward a photonic demonstration of device-independent quantum key
  distribution''.
\newblock \href{https://dx.doi.org/10.1103/PhysRevLett.129.050502}{Phys. Rev.
  Lett. {\bf 129}, 050502}~(2022).

\bibitem{Gilchrist1999}
A.~Gilchrist, P.~Deuar, and M.~D. Reid.
\newblock ``Contradiction of quantum mechanics with local hidden variables for
  quadrature phase measurements on pair-coherent states and squeezed
  macroscopic superpositions of coherent states''.
\newblock \href{https://dx.doi.org/10.1103/PhysRevA.60.4259}{Phys. Rev. A {\bf
  60}, 4259--4271}~(1999).

\bibitem{Munro1999}
W.~J. Munro.
\newblock ``Optimal states for {B}ell-inequality violations using
  quadrature-phase homodyne measurements''.
\newblock \href{https://dx.doi.org/10.1103/PhysRevA.59.4197}{Phys. Rev. A {\bf
  59}, 4197--4201}~(1999).

\bibitem{Wenger2003}
J\'er\^ome Wenger, Mohammad Hafezi, Fr\'ed\'eric Grosshans, Rosa Tualle-Brouri,
  and Philippe Grangier.
\newblock ``Maximal violation of {B}ell inequalities using continuous-variable
  measurements''.
\newblock \href{https://dx.doi.org/10.1103/PhysRevA.67.012105}{Phys. Rev. A
  {\bf 67}, 012105}~(2003).

\bibitem{Oudot2024}
Enky Oudot, Gaël Massé, Xavier Valcarce, and Antonio Acín.
\newblock ``Realistic {B}ell tests with homodyne measurements''~(2024).
\newblock  \href{http://arxiv.org/abs/2402.01530}{arXiv:2402.01530}.

\bibitem{Garcia2004}
R.~Garc\'{\i}a-Patr\'on, J.~Fiur\'a\ifmmode~\check{s}\else \v{s}\fi{}ek, N.~J.
  Cerf, J.~Wenger, R.~Tualle-Brouri, and Ph. Grangier.
\newblock ``Proposal for a loophole-free {B}ell test using homodyne
  detection''.
\newblock \href{https://dx.doi.org/10.1103/PhysRevLett.93.130409}{Phys. Rev.
  Lett. {\bf 93}, 130409}~(2004).

\bibitem{Garcia2005}
Ra\'ul Garc\'{\i}a-Patr\'on, Jarom\'{\i}r Fiur\'a\ifmmode~\check{s}\else
  \v{s}\fi{}ek, and Nicolas~J. Cerf.
\newblock ``Loophole-free test of quantum nonlocality using high-efficiency
  homodyne detectors''.
\newblock \href{https://dx.doi.org/10.1103/PhysRevA.71.022105}{Phys. Rev. A
  {\bf 71}, 022105}~(2005).

\bibitem{Valcarce2020}
Xavier Valcarce, Pavel Sekatski, Davide Orsucci, Enky Oudot, Jean-Daniel
  Bancal, and Nicolas Sangouard.
\newblock ``What is the minimum {CHSH} score certifying that a state resembles
  the singlet?''.
\newblock \href{https://dx.doi.org/10.22331/q-2020-03-23-246}{Quantum {\bf 4},
  246}~(2020).

\bibitem{Kaniewski2016}
Jędrzej Kaniewski.
\newblock ``Analytic and nearly optimal self-testing bounds for the
  {C}lauser-{H}orne-{S}himony-{H}olt and {M}ermin inequalities''.
\newblock \href{https://dx.doi.org/10.1103/physrevlett.117.070402}{Physical
  Review Letters {\bf 117}, 070402}~(2016).

\bibitem{Tanzilli2012}
S.~Tanzilli, A.~Martin, F.~Kaiser, M.P.~De Micheli, O.~Alibart, and D.B.
  Ostrowsky.
\newblock ``On the genesis and evolution of integrated quantum optics''.
\newblock \href{https://dx.doi.org/10.1002/lpor.201100010}{Laser {\&} Photonics
  Reviews {\bf 6}, 115--143}~(2012).

\bibitem{Pelucchi2021}
Emanuele Pelucchi, Giorgos Fagas, Igor Aharonovich, Dirk Englund, Eden
  Figueroa, Qihuang Gong, H\"{u}bel Hannes, Jin Liu, Chao-Yang Lu, Nobuyuki
  Matsuda, Jian-Wei Pan, Florian Schreck, Fabio Sciarrino, Christine
  Silberhorn, Jianwei Wang, and Klaus~D. J\"{o}ns.
\newblock ``The potential and global outlook of integrated photonics for
  quantum technologies''.
\newblock \href{https://dx.doi.org/10.1038/s42254-021-00398-z}{Nature Reviews
  Physics {\bf 4}, 194--208}~(2021).

\bibitem{Mnih2015}
Volodymyr Mnih, Koray Kavukcuoglu, David Silver, Andrei~A. Rusu, Joel Veness,
  Marc~G. Bellemare, Alex Graves, Martin Riedmiller, Andreas~K. Fidjeland,
  Georg Ostrovski, Stig Petersen, Charles Beattie, Amir Sadik, Ioannis
  Antonoglou, Helen King, Dharshan Kumaran, Daan Wierstra, Shane Legg, and
  Demis Hassabis.
\newblock ``Human-level control through deep reinforcement learning''.
\newblock \href{https://dx.doi.org/10.1038/nature14236}{Nature {\bf 518},
  529–533}~(2015).

\bibitem{Mnih2016}
Volodymyr Mnih, Adri\`{a}~Puigdom\`{e}nech Badia, Mehdi Mirza, Alex Graves, Tim
  Harley, Timothy~P. Lillicrap, David Silver, and Koray Kavukcuoglu.
\newblock ``Asynchronous methods for deep reinforcement learning''.
\newblock In Proceedings of the 33rd International Conference on International
  Conference on Machine Learning - Volume 48.
\newblock \href{https://dx.doi.org/10.48550/arXiv.1602.01783}{Page
  1928–1937}.
\newblock ICML'16. JMLR.org~(2016).

\bibitem{Haarnoja2018}
Tuomas Haarnoja, Aurick Zhou, Kristian Hartikainen, George Tucker, Sehoon Ha,
  Jie Tan, Vikash Kumar, Henry Zhu, Abhishek Gupta, Pieter Abbeel, and Sergey
  Levine.
\newblock ``Soft actor-critic algorithms and applications''~(2018).
\newblock  \href{http://arxiv.org/abs/1812.05905}{arXiv:1812.05905}.

\bibitem{Schulman2015}
John Schulman, Sergey Levine, Philipp Moritz, Michael~I. Jordan, and Pieter
  Abbeel.
\newblock ``Trust region policy optimization''~(2015).
\newblock  \href{http://arxiv.org/abs/1502.05477}{arXiv:1502.05477}.

\bibitem{Schulman2017}
John Schulman, Filip Wolski, Prafulla Dhariwal, Alec Radford, and Oleg Klimov.
\newblock ``Proximal policy optimization algorithms''~(2017).
\newblock  \href{http://arxiv.org/abs/1707.06347}{arXiv:1707.06347}.

\bibitem{Carleo2019}
Giuseppe Carleo, Ignacio Cirac, Kyle Cranmer, Laurent Daudet, Maria Schuld,
  Naftali Tishby, Leslie Vogt-Maranto, and Lenka Zdeborov\'a.
\newblock ``Machine learning and the physical sciences''.
\newblock \href{https://dx.doi.org/10.1103/RevModPhys.91.045002}{Rev. Mod.
  Phys. {\bf 91}, 045002}~(2019).

\bibitem{Biamonte2017}
Jacob Biamonte, Peter Wittek, Nicola Pancotti, Patrick Rebentrost, Nathan
  Wiebe, and Seth Lloyd.
\newblock ``Quantum machine learning''.
\newblock \href{https://dx.doi.org/10.1038/nature23474}{Nature {\bf 549},
  195--202}~(2017).

\bibitem{Dunjko2018}
Vedran Dunjko and Hans~J. Briegel.
\newblock ``Machine learning \& artificial intelligence in the quantum domain:
  a review of recent progress''.
\newblock \href{https://dx.doi.org/10.1088/1361-6633/aab406}{Rep. Prog. Phys.
  {\bf 81}, 074001}~(2018).

\bibitem{Krenn2020}
Mario Krenn, Manuel Erhard, and Anton Zeilinger.
\newblock ``Computer-inspired quantum experiments''.
\newblock \href{https://dx.doi.org/10.1038/s42254-020-0230-4}{Nature Reviews
  Physics {\bf 2}, 649--661}~(2020).

\bibitem{Krenn2016}
Mario Krenn, Mehul Malik, Robert Fickler, Radek Lapkiewicz, and Anton
  Zeilinger.
\newblock ``Automated search for new quantum experiments''.
\newblock \href{https://dx.doi.org/10.1103/physrevlett.116.090405}{Physical
  Review Letters {\bf 116}, 090405}~(2016).

\bibitem{Melnikov2020}
Alexey~A. Melnikov, Pavel Sekatski, and Nicolas Sangouard.
\newblock ``Setting up experimental {B}ell tests with reinforcement learning''.
\newblock \href{https://dx.doi.org/10.1103/PhysRevLett.125.160401}{Phys. Rev.
  Lett. {\bf 125}, 160401}~(2020).

\bibitem{Krenn2021}
Mario Krenn, Jakob~S. Kottmann, Nora Tischler, and Al{\'{a}}n Aspuru-Guzik.
\newblock ``Conceptual understanding through efficient automated design of
  quantum optical experiments''.
\newblock \href{https://dx.doi.org/10.1103/physrevx.11.031044}{Physical Review
  X {\bf 11}, 031044}~(2021).

\bibitem{Valcarce2023}
X.~Valcarce, P.~Sekatski, E.~Gouzien, A.~Melnikov, and N.~Sangouard.
\newblock ``Automated design of quantum-optical experiments for
  device-independent quantum key distribution''.
\newblock \href{https://dx.doi.org/10.1103/PhysRevA.107.062607}{Phys. Rev. A
  {\bf 107}, 062607}~(2023).

\bibitem{Tsirelson1987}
B.~S. Tsirel'son.
\newblock ``Quantum analogues of the bell inequalities. the case of two
  spatially separated domains''.
\newblock \href{https://dx.doi.org/10.1007/BF01663472}{Journal of Soviet
  Mathematics {\bf 36}, 557--570}~(1987).

\bibitem{Bell1986}
John~S. Bell.
\newblock ``{EPR} correlations and {EPW} distributions''.
\newblock
  \href{https://dx.doi.org/https://doi.org/10.1111/j.1749-6632.1986.tb12429.x}{Annals
  of the New York Academy of Sciences {\bf 480}, 263--266}~(1986).

\bibitem{Jabbour2023}
Michael~G. Jabbour and Jonatan~Bohr Brask.
\newblock ``Constructing local models for general measurements on bosonic
  {G}aussian states''.
\newblock \href{https://dx.doi.org/10.1103/PhysRevLett.131.110202}{Phys. Rev.
  Lett. {\bf 131}, 110202}~(2023).

\bibitem{Kuzmich2000}
A.~Kuzmich, I.~A. Walmsley, and L.~Mandel.
\newblock ``Violation of {B}ell's inequality by a generalized
  {E}instein-{P}odolsky-{R}osen state using homodyne detection''.
\newblock \href{https://dx.doi.org/10.1103/PhysRevLett.85.1349}{Phys. Rev.
  Lett. {\bf 85}, 1349--1353}~(2000).

\bibitem{Walschaers2021}
Mattia Walschaers.
\newblock ``Non-{G}aussian quantum states and where to find them''.
\newblock \href{https://dx.doi.org/10.1103/PRXQuantum.2.030204}{PRX Quantum
  {\bf 2}, 030204}~(2021).

\bibitem{code_repo}
Corentin Lanore, Federico Grasselli, and Xavier Valcarce.
\newblock
  code:~\href{https://github.com/JuliaReinforcementLearning/ReinforcementLearning.jl}{JuliaReinforcementLearning/ReinforcementLearning.jl}.

\bibitem{Valcarce2021}
Xavier Valcarce.
\newblock
  code:~\href{https://github.com/xvalcarce/QuantumOpticalCircuits.jl}{xvalcarce/QuantumOpticalCircuits.jl}.

\bibitem{Tian2020}
Jun Tian and other contributors.
\newblock
  code:~\href{https://github.com/JuliaReinforcementLearning/ReinforcementLearning.jl}{JuliaReinforcementLearning/ReinforcementLearning.jl}.

\bibitem{Vogel2006}
Werner Vogel and Dirk‐Gunnar Welsch.
\newblock ``Quantum optics''.
\newblock \href{https://dx.doi.org/10.1002/3527608524}{Wiley}. ~(2006).

\bibitem{Pirandola2009}
Stefano Pirandola, Alessio Serafini, and Seth Lloyd.
\newblock ``Correlation matrices of two-mode bosonic systems''.
\newblock \href{https://dx.doi.org/10.1103/PhysRevA.79.052327}{Phys. Rev. A
  {\bf 79}, 052327}~(2009).

\bibitem{JuliaRLcommit}
Jun Tian and other contributors~(2020).
\newblock
  code:~\href{https://github.com/JuliaReinforcementLearning/ReinforcementLearning.jl/blob/b4c9f404be5b921178fbee34f93a00dd143a829a/src/ReinforcementLearningZoo/src/algorithms/policy_gradient/ppo.jl}{JuliaReinforcementLearning/ReinforcementLearning.jl
  commit:b4c9f40}.

\bibitem{Brunner2007}
Nicolas Brunner, Nicolas Gisin, Valerio Scarani, and Christoph Simon.
\newblock ``Detection loophole in asymmetric bell experiments''.
\newblock \href{https://dx.doi.org/10.1103/PhysRevLett.98.220403}{Phys. Rev.
  Lett. {\bf 98}, 220403}~(2007).

\bibitem{Cabello2007}
Ad\'an Cabello and Jan-\AA{}ke Larsson.
\newblock ``Minimum detection efficiency for a loophole-free atom-photon bell
  experiment''.
\newblock \href{https://dx.doi.org/10.1103/PhysRevLett.98.220402}{Phys. Rev.
  Lett. {\bf 98}, 220402}~(2007).

\end{thebibliography}

\appendix

\onecolumn

\section{Photonic circuits}
\label{app:quantopt}

\subsection{Gaussian states}
\label{sec:gaussianstates}

We consider circuits composed of $N$ bosonic modes. Each mode is identified with an index  $l\in\{1,\dots,N\}$ and is associated with the ladder operators $\hat{a}_l$ and $\hat{a}_l^\dagger$ or, alternatively, to the dimensionless quadrature field operators $\hat{x}_l = \frac{\hat{a}_l^\dagger + \hat{a}_l}{2}$ and $\hat{p}_l = i\frac{\hat{a}_l^\dagger - \hat{a}_l}{2}$, which satisfy the commutation relation: $[\hat{x}_l,\hat{p}_l]=i\frac{1}{2}$. By arranging the quadrature field operators in the vector:  $\mathbf{\hat{q}} = (\hat x_1, \hat p_1,\ldots, \hat x_N, \hat p_N)$, we have the commutation relation
\begin{align}
    [\hat{q}_k,\hat{q}_j] = i \frac{\Omega_{k,j}}{2} \label{commrel},
\end{align}
with $\hat{q}_k$ the $k$-th component of the vector $\mathbf{\hat{q}}$ and $\Omega$ the symplectic matrix
\begin{align}
    \Omega = \bigoplus_{l=1}^N \omega, \quad \omega=  \begin{pmatrix}
    0 & 1  \\
    -1 & 0
    \end{pmatrix}.
\end{align}

The Wigner function of the state $\rho$ of $N$ bosonic modes is a quasi-probability distribution~\cite{Vogel2006} defined as
\begin{equation}
		\label{eq:Wignerfunction}
        W_\rho(\tilde{\balpha}) = \tr \left( \, \hat{\rho} \, \bigotimes_{l=1}^N \delta(\hat{a}_l-\alpha_l ) \right), 
\end{equation}
where $\tilde{\balpha}\in\mathds{R}^{2N}$ is the vector $\tilde{\balpha} = \{\text{Re}{(\alpha_1)}, \text{Im}{(\alpha_1)}, \hdots, \text{Re}{(\alpha_n)}, \text{Im}{(\alpha_n})\}^T$, and
\begin{align}
    \delta(\hat{a}-\alpha )= \frac{1}{\pi^2} \int \text{d}^2 \beta \,\, e^{(\hat{a}^\dag-\alpha^*)\beta - (\hat{a}-\alpha)\beta^*},
\end{align}
where $\text{d}^2 \beta=\text{d} \text{Re}(\beta) \, \text{d} \text{Im}(\beta)$. 
A quantum state on $N$ bosonic modes can be expressed in terms of its Wigner function following
\begin{equation}
 \hat{\rho} = \pi^N \int d^{2N}\tilde{\balpha}\, W(\tilde{\balpha})\,   \bigotimes_{l=1}^N \delta(\hat{a}_l-\alpha_l ).
\end{equation}
$\hat{\rho}$ is a Gaussian state if its Wigner function is equal to the probability density function of a multivariate normal distribution. That is, if it can be expressed in the form
\begin{align}
	\label{eq:AnnexWigner}
	W_{{\bm \mu},\Sigma} (\tilde{\balpha}) &= \frac{\exp\left[-\frac{1}{2}(\tilde\balpha-\bmu)^T\Sigma^{-1}(\tilde\balpha-\bmu)\right]}{(2\pi)^n\sqrt{\det\Sigma}},
\end{align}
where the displacement vector $\bmu$ and the covariance matrix $\Sigma$ have for elements
\begin{align}
	\mu_j &= \mean{\hat{q}_j}, \\
    \Sigma_{j,k} &= \frac{1}{2}\mean{\hat{q_j} \hat{q_k} + \hat{q_k} \hat{q_j}} - \mu_j \mu_k.
\end{align}

Note that, due to the commutation relations in \cref{commrel}, an arbitrary real symmetric matrix $\Sigma$ is a valid covariance matrix if and only if the inequality
\begin{align}
    \Sigma + i \frac{\Omega}{4} \geq 0.
\end{align}
is satisfied~\cite{Pirandola2009}.

The total number of parameters for an $N$-mode Gaussian state consists of $2N(2N+1)$ elements from the upper triangular part of the covariance matrix $\Sigma$ (accounting for its symmetry), along with $2N$ elements from the displacement vector.

\subsection{Gaussian operations} 
\label{sec:GaussianOperations}

Gaussian operations are particular unitary transformations $\hat{U} = e^{i\hat{H}}$ which are generated by Hamiltonians $\hat{H}$ that are linear and bilinear in the field modes. Such transformations are called Gaussian in that they preserve the Gaussian character of the state. As a matter of fact, the transformation induced on the quadrature operators preserves the commutation relations and reads
\begin{align}
    U^\dag \mathbf{\hat{q}} U = M \mathbf{\hat{q}} + \bm{d},
\end{align}
where $\bm{d}$ is a real vector and $M$ is a symplectic transformation, satisfying $M \Omega M^T = \Omega$, which characterize the Gaussian operation. Hence, given a Gaussian state with covariance matrix $\Sigma$ and displacement vector $\bmu$, the resulting Gaussian state after applying a Gaussian operation $(\bm{d},M)$ is described by
\begin{align}
    \bmu' &= M \bmu + \bm{d}, \\
    \Sigma'&= M \Sigma M^T.
\end{align}

In \cref{table:GaussianOperations} we report the Gaussian operations, also named gates, that are considered in this manuscript, in terms of both the unitary transformation and the corresponding vector $\bm{d}$ and symplectic matrix $M$. We adopt the notation for which the square brackets $[j,k]$ indicate that the gate acts non-trivially on modes $j$ and $k$, with $j<k$. Note that we restrict the parameters of each operation to the real domain, as any Gaussian operation with a complex parameter can be achieved with a combination of Gaussian operations with real parameters and phase shifters.

\begin{table}[!h]
\begin{tabular}{ c l l}
\toprule
Gaussian Operation & Unitary Operator & Symplectic Matrix \\
\midrule
Phase Shifter & $\hat{R}(\theta)[l] = e^{-i\theta\hat{a}^{\dagger}_l\hat{a}_l}$ & $  M= \left(\begin{array}{c|c|c}
\mathbb{I} & 0 & 0 \\
\hline
0 & \begin{matrix}
    \cos(\theta) & \sin(\theta) \\
    -\sin(\theta) & \cos(\theta) \\
    \end{matrix} & 0 \\
\hline
0 & 0 & \mathbb{I}
\end{array}\right)$ \\
\midrule
Single Mode Squeezer & $\hat{S}_1(r)[l] =e^{r[\hat{a}_l^2-(\hat{a}_l^{\dagger})^2]/2}$ & $M= \left(\begin{array}{c|c|c}
\mathbb{I} & 0 & 0 \\
\hline
0 & \begin{matrix}
    e^{-r} & 0 \\
    0 & e^{r} \\
    \end{matrix} & 0 \\
\hline
0 & 0 & \mathbb{I}
\end{array}\right)$ \\
\midrule
Beam Splitter & $\hat{B}(\theta)[j,k] = e^{\theta(\hat{a}_j^{\dagger}\hat{a}_k-\hat{a}_j\hat{a}_k^{\dagger})  }$ & $ M=\left(\begin{array}{c|c|c|c|c}
  \mathbb{I} & 0 & 0 & 0 & 0 \\
  \hline
  0 & \cos(\theta) \,\mathbb{I}_2 & 0 & \sin(\theta) \,\mathbb{I}_2 & 0 \\
  \hline
  0 & 0 & \mathbb{I} & 0 & 0 \\
  \hline
  0 & -\sin(\theta) \,\mathbb{I}_2 & 0 & \cos(\theta) \,\mathbb{I}_2 & 0 \\
  \hline
  0 & 0 & 0 & 0 & \mathbb{I}
\end{array}\right)$ \\

\midrule

Two Modes Squeezer & $\hat{S}_2(r)[j,k] = e^{r(\hat{a}_j\hat{a}_k-\hat{a}_j^{\dagger}\hat{a}_k^{\dagger})}$ & $ M =   \left(\begin{array}{c|c|c|c|c}
  \mathbb{I} & 0 & 0 & 0 & 0 \\
  \hline
  0 & \cosh(r) \,\mathbb{I}_2 & 0 & -\sinh(r) \,\mathbb{I}_2 & 0 \\
  \hline
  0 & 0 & \mathbb{I} & 0 & 0 \\
  \hline
  0 & -\sinh(r) \,\mathbb{I}_2 & 0 & \cosh(r) \,\mathbb{I}_2 & 0 \\
  \hline
  0 & 0 & 0 & 0 & \mathbb{I}
\end{array}\right)$\\
\bottomrule

\end{tabular}
\caption{The Gaussian operations considered in this manuscript, when acting on an $N$-mode Gaussian state. Here $\mathbb{I}_n$ is the $n \times n$ identity matrix and $\sigma_z$ is the third Pauli matrix. Note that the non-trivial $2 \times 2$ blocks in the symplectic matrices correspond to the modes $j$ and $k$, i.e.~to rows/columns pairs $2j-1,2j$ and $2k-1,2k$. For all the operations considered, $\bm{d}=0$.}\vspace{1ex}
\label{table:GaussianOperations}
\end{table}

Importantly, Euler's decomposition ensures that every symplectic transformation $M$ can be decomposed in the form
\begin{align}
    M = O \begin{pmatrix}
    D & 0 \\
    0 & D^{-1}
    \end{pmatrix}
    O' \label{euler-decomp}, 
\end{align}
where $O$ and $O'$ are orthogonal symplectic matrices and $D$ is diagonal and positive definite. Physically, this implies that every Gaussian operation can be implemented by a sequence of passive beam-splitter and phase shifter gates (described by the orthogonal matrix $O'$), followed by squeezer gates (matrix $D$) and more passive gates (matrix $O$). Moreover, if the initial Gaussian state is the vacuum -- as is in our setup --, the first sequence of passive gates can be omitted since it acts
trivially on the vacuum. Therefore, in our case, any Gaussian operation can be attained by an array of single mode squeezers, one per mode, followed by passive gates.

\subsection{Photon heralding}
\label{sec:heralding}

In this subsection, we review the conditional quantum state obtained by heralding the presence or absence of photons in a subset of its modes. The derivation of the formulas presented here can be found in~\cite{Valcarce2023}.

Consider an $N$-mode Gaussian state $\hat{\rho}$ with covariance matrix $\Sigma$ and displacement vector $\bmu$. The quantum state $\hat{\rho}_{\neg l}=\Tr_l[\hat{\rho}]$ obtained by tracing out the $l$-th mode is still Gaussian, with displacement vector $\bmu_{\neg l} = \mathrm{TR}_l (\bmu)$ and covariance matrix $\Sigma_{\neg l}= \mathrm{TR}_l (\Sigma)$, where we defined $\mathrm{TR}_l (\cdot)$ as the operation that removes the rows and/or columns of a vector or matrix at positions $2l-1$ and $2l$.

Consider now the quantum state $\hat{\rho}_{\circ l}$ obtained from $\hat{\rho}$ by measuring the $l$-th mode with a threshold detector, with efficiency $\eta$, and conditioning on a no-click event. By definition, the corresponding unnormalized state is $p_{\circ l} \hat{\rho}_{\circ l} = \Tr_l[(1-\eta)^{\hat{a}^\dag_l \hat{a}_l} \hat{\rho}]$. One can show that the conditional state $\hat{\rho}_{\circ l}$ is still Gaussian, with displacement vector and covariance matrix given by, respectively,
\begin{align}
    \bmu_{\circ l} &= \mathrm{TR}_l[(\Sigma^{-1} + F)^{-1} \Sigma^{-1} \bmu] \nonumber \\
    \Sigma_{\circ l} &= \mathrm{TR}_l[(\Sigma^{-1} + F)^{-1} ] , \label{state-heraldnoclick}
\end{align}
where the $2N \times 2N$ matrix $F$ is null everywhere except for the elements at position $2l-1$ and $2l$, which we write
\begin{align}
    F= \frac{4 \eta}{2-\eta} \left(\begin{array}{c|c|c}
0 & 0 & 0 \\
\hline
0 &  \mathbb{I}_2 & 0 \\
\hline
0 & 0 & 0
\end{array}\right).
\end{align}
Moreover, the probability of the no-click event is given by
\begin{align}
    p_{\circ l} = \frac{2}{2-\eta} \sqrt{\frac{(\det \Sigma)^{-1}}{\det(\Sigma^{-1} + F)}} \, e^{-\frac{1}{2}\bmu^T \left[\Sigma^{-1} - \Sigma^{-1} (\Sigma^{-1}+F)^{-1} \Sigma^{-1} \right]\bmu} . \label{herald-prob-noclick}
\end{align}

Consider now  the quantum state $\hat{\rho}_{\bullet l}$ obtained from $\hat{\rho}$ by measuring the $l$-th mode with a threshold detector, with efficiency $\eta$, and conditioning on a click event. By definition, the corresponding unnormalized state is $p_{\bullet l} \hat{\rho}_{\bullet l} = \Tr_l[(\mathbb{I}-(1-\eta)^{\hat{a}^\dag_l \hat{a}_l}) \hat{\rho}]$. We deduce that $\hat{\rho}_{\bullet l}$ is a linear combination of two Gaussian states with a negative coefficient
\begin{align}
    \hat{\rho}_{\bullet l} = \frac{1}{p_{\bullet l}} \hat{\rho}_{\neg l}  - \frac{p_{\circ l}}{p_{\bullet l}} \hat{\rho}_{\circ l}, \label{state-heraldclick}
\end{align}
where the heralding probability is $p_{\bullet l} = 1 -p_{\circ l}$, with $p_{\circ l}$ defined in \cref{herald-prob-noclick}.

\subsection{CHSH score from homodyne measurements}
\label{app:chsh_computation}

In order to numerically compute the correlators $\braket{A_x B_y}$ appearing in the CHSH score, we start by expressing the joint probability of Alice's and Bob's quadrature outcomes as a linear combination of Gaussian distributions. In the following, we omit the explicit indication of Alice's and Bob's measurement inputs for ease of notation. We obtain
\begin{align}
    P(x_1 , x_2) &= \iint_{-\infty}^{\infty} \text{d}p_1 \text{d}p_2 \tilde{W}_{12} (x_1,p_1,x_2,p_2) \\
    &= \sum_k w_k \iint_{-\infty}^{\infty}  \text{d}p_1 \text{d}p_2 W^k_{12} (x_1,p_1,x_2,p_2) \\
    &= \sum_k w_k \frac{\exp\left[-\frac{1}{2}(\tilde\balpha-\bmu_k)^T\sigma^{-1}_k(\tilde\balpha-\bmu_k)\right]}{2\pi\sqrt{\det\sigma_k}}, \label{joint-prob}
\end{align}
where in the first equality we express the Wigner function of the heralded state as a linear combination of Gaussian Wigner functions, $W^k_{12}$, according to \cref{pseudo-gaussian-states}, while in the second equality we compute the marginal of a Gaussian distribution. In particular, the $2 \times 2$ matrix $\sigma_k$ is obtained from the covariance matrix of the $k$-th Guassian state by removing the second and fourth row and column, while the 2-dimensional vector $\bmu_k$ is obtained from the first and third entry of the displacement vector of the $k$-th Gaussian state. Finally, we have that $\tilde{\balpha}^T=(x_1,x_2)$.

Although the matrices $\sigma_k$ are symmetric and satisfy $\sigma_k \geq 0$, the exponential functions in \cref{joint-prob} are not necessarily Gaussian. This is because integrating a Gaussian Wigner function over the variables $p_1$ and $p_2$ does not always result in a Gaussian distribution. In particular, if the matrices $\sigma_k$ are singular, the expression becomes undefined. However, in practice, the determinants are never exactly zero, allowing us to safely use the expression in
\cref{joint-prob}.

Now, let us consider a general binning function $a(x_1):\mathbb{R} \to \{-1,1\}$ for Alice and $b(x_2):\mathbb{R} \to \{-1,1\}$ for Bob. The special case of sign binning is recovered for $a(x)=b(x)=1$ for $x>0$ and $a(x)=b(x)=-1$ otherwise. Then, the generic correlator appearing in the CHSH score is expressed in terms of the joint probability $P(x_1,x_2)$ as
\begin{align}
    \braket{A_x B_y} &= \iint_{-\infty}^\infty \text{d}x_1 \text{d}x_2 \, a(x_1) b(x_2) P(x_1, x_2).
\end{align}
By employing the result in \cref{joint-prob}, we can recast the above expression as
\begin{equation}
	\begin{split}
		\braket{A_x B_y} = \sum_k w_k & \left[ \iint_{R_+}\text{d}x_1 \text{d}x_2  \frac{\exp\left[-\frac{1}{2}(\tilde\balpha-\bmu_k)^T\sigma^{-1}_k(\tilde\balpha-\bmu_k)\right]}{2\pi\sqrt{\det\sigma_k}}\right. \\
					 &\left. - \iint_{R_-} \text{d}x_1 \text{d}x_2 \frac{\exp\left[-\frac{1}{2}(\tilde\balpha-\bmu_k)^T\sigma^{-1}_k(\tilde\balpha-\bmu_k)\right]}{2\pi\sqrt{\det\sigma_k}} \right],
	\end{split}
\end{equation}
where we defined regions $R_\pm$, $R_\pm = \left\lbrace(x_1,x_2) : a(x_1)=\pm b(x_2) \right\rbrace$, such that $\mathbb{R}^2=R_+ \cup R_-$. Considering that the integrand function is normalized to one and that $\sum_k w_k=1$, we can write
\begin{align}
    \braket{A_x B_y} &= -1 + 2\sum_k w_k \iint_{R_+}\text{d}x_1 \text{d}x_2  \frac{\exp\left[-\frac{1}{2}(\tilde\balpha-\bmu_k)^T\sigma^{-1}_k(\tilde\balpha-\bmu_k)\right]}{2\pi\sqrt{\det\sigma_k}}. 
	\label{correlator-calc}
\end{align}

Note that the region $R_+$ is given by a set of disjoint rectangular areas $R_{\square}=[u_1,v_1]\times [u_2,v_2]$ that depend on the specific choice of binning functions $a(x)$ and $b(x)$. Thus, for the sake of calculating \cref{correlator-calc}, we are interested in the integral
\begin{align}
    I = \int_{u_1}^{v_1}\text{d}x_1 \int_{u_2}^{v_2}\text{d}x_2 \exp\left[-\frac{1}{2}\tilde\balpha^T A \tilde\balpha\right], \label{Int}
\end{align}
where $A$ is a symmetric positive semi-definite matrix
\begin{align}
    A = \begin{pmatrix}
    r & t \\
    t & s
    \end{pmatrix}.
\end{align}
By partially integrating over $x_1$, one can show that the integral in \cref{Int} reduces to
\begin{align}
    I &= \int_{\frac{u_2}{\sqrt{2}}}^{\frac{v_2}{\sqrt{2}}} \text{d}x_2 \, e^{-x_2^2\frac{rs-t^2}{r}} \sqrt{\frac{\pi}{r}} \left[\mathrm{Erf}\left(\sqrt{\frac{r}{2}} v_1 + \frac{t}{\sqrt{r}} x_2\right) - \mathrm{Erf}\left(\sqrt{\frac{r}{2}} u_1 + \frac{t}{\sqrt{r}} x_2\right)\right], \label{Int2}
\end{align}
where we introduced the error function
\begin{align}
    \mathrm{Erf}(z) = \frac{2}{\sqrt{\pi}} \int_0^z \text{d}t \, e^{-t^2}. 
\end{align}

We recast the correlator in \cref{correlator-calc} as 
\begin{align}
    \braket{A_x B_y} &= -1 + 2\sum_k w_k \sum_{R_{\square}\in R_+} I_{\square} \label{correlator-calc2},
\end{align}
where the quantities $I_{\square}$ are obtained via \cref{Int2} and read
\begin{align}
    I_{\square} &= \int_{u_1}^{v_1}\text{d}x_1 \int_{u_2}^{v_2}\text{d}x_2 \frac{\exp\left[-\frac{1}{2}(\tilde\balpha-\bmu_k)^T\sigma^{-1}_k(\tilde\balpha-\bmu_k)\right]}{2\pi\sqrt{\det\sigma_k}} \\
    &= \frac{1}{2\sqrt{\pi}} \int_{\frac{u_2-(\bmu_k)_2}{\sqrt{2\sigma_b}}}^{\frac{v_2-(\bmu_k)_2}{\sqrt{2\sigma_b}}} \text{d} y e^{-y^2} \left[\mathrm{Erf}\left(\frac{\sqrt{\sigma_b}\frac{v_1-(\bmu_k)_1}{\sqrt{2}}-\sigma_c y}{\sqrt{\det\sigma_k}}\right) - \mathrm{Erf}\left(\frac{\sqrt{\sigma_b}\frac{u_1-(\bmu_k)_1}{\sqrt{2}}-\sigma_c y}{\sqrt{\det\sigma_k}}\right)\right], \label{Intrect}
\end{align}
where we parametrized the matrix $\sigma_k$ as
\begin{align}
    \sigma_k = \begin{pmatrix}
    \sigma_a & \sigma_c \\
    \sigma_c & \sigma_b
    \end{pmatrix}.
\end{align}

One can then employ \cref{correlator-calc2}, in combination with a numerical computation of \cref{Intrect}, to compute the CHSH score for any arbitrary choice of binning functions. For the specific choice that we made in our work, i.e.~sign binning, we have
\begin{align}
    \braket{A_x B_y} &= -1 + 2\sum_k w_k (I_k + I'_k) \label{correlator-signbin},
\end{align}
where $I_k$ and $I'_k$ are obtained by numerically solving \cref{Intrect} with parameters $u_1=u_2=0$ and $v_1=v_2=\infty$ for $I_k$ and $u_1=u_2=-\infty$ and $v_1=v_2=0$ for $I'_k$.

\section{Automated generation of photonic Bell tests} \label{app:circuits}

\subsection{Exploring photonic setups}

We explore $N$-mode photonic circuits composed of $n_\mathrm{circuit}$ optical elements.
Photonic operations are applied to the $N$-modes during a state preparation phase.
The last $N-2$ modes are then heralded, while the first two modes are sent to Alice and Bob respectively.

To enhance the experimental relevance of discovered circuit and the chance to find interesting setups, we devise exploration \emph{strategies} ; constraints on the initial state and on the type of the available photonic operations.
The strategies are detailed in \cref{tab:strategies}.
These strategies are physically-informed, they leverage knowledge on photonic state construction and answer practical experimental requirements.

We remark that all strategies but Strategy 1 are characterized by having the initial vacuum modes initialized by active gates $S_1$ and $S_2$, followed by only passive gates. This approach has some advantages. First of all, by virtue of Euler's decomposition in \cref{euler-decomp}, this approach is non-restrictive as it still enables us to explore the whole set of circuits that can be generated by the full set of gates. 
Second, we avoid circuits with multiple squeezers in sequence, which not only may be experimentally challenging but can also cause numerical errors due to a compounded large squeezing parameter, thereby invalidating the results.

\begin{table}
	\begin{center}
		\begin{tabular}[c]{c | l | l}
			\hline
			\multicolumn{1}{c|}{\textbf{Strategy}} &  \multicolumn{1}{c|}{\textbf{Initial state}} & \multicolumn{1}{c}{}{\textbf{ Photonic operations}} \\
			\hline
			1 & Vacuum state & $S_1$,$S_2$,$R$,$B$  \\
			2 & $S_2$ gate on mode 1,2 & Passive gates ($R$,$B$) \\
			3 & $S_2$ gate every pair of modes (1,2~/~3,4~/~..) & Passive gates ($R$,$B$) \\
			4 & $S_1$ gate on mode 1 and 2 & Passive gates ($R$,$B$) \\
			5 & $S_1$ gate on every mode & Passive gates ($R$,$B$) \\
			\hline
		\end{tabular}
		\caption{Circuit exploration strategies. A strategy is characterized by a fixed initial state construction followed by specific photonic operations. $S_1$ and $S_2$ label single- and two-mode squeezer, $R$ is phase-shifter, and $B$ represents beam-splitter.  }\label{tab:strategies}
	\end{center}
\end{table}

\begin{figure}[b]
	\begin{center}
  \subfloat[Heralding on click.]{
	  \label{fig:heralding1}%
 	  \includegraphics[width=.25\linewidth]{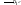}
  }\hfill
  \subfloat[Approximating single-photon projection.]{
	  \label{fig:heralding2}%
	  \includegraphics[width=.4\linewidth]{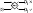}
  }
\caption{Two alternative ways to herald a mode. In (a), we herald the state on a click in a threshold detector. In (b), we approximate a projection on a single photon by mixing the mode with the vacuum through a beam-splitter with parameter $\theta=0.1$ (the transmittance of the beam-splitter is $\cos(0.1)\approx 0.995$). Then, we herald on a click in the threshold detector at the reflected port and a no-click in the transmitted port. This combined event heralds on the detection of a single photon with high probability.}
\label{fig:heralding}
\end{center}
\end{figure}

For each strategy, we run two independent search for each of two heralding mechanisms.
In Fig.~\ref{fig:heralding} we schematize these two mechanisms on a single mode. 
One possibility is to herald the state on a click in a single threshold detector, as depicted in \cref{fig:heralding}(a) (c.f. 
Eq.~\eqref{state-heraldclick} in \cref{app:quantopt}). 
Alternatively, we approximate the measurement that projects on a single photon, namely the POVM $\{\ket{1}\bra{1}, \mathbb{I} - \ket{1}\bra{1}\}$, where $\ket{1}$ is a Fock state, with the setup depicted in \cref{fig:heralding}(b). 
We achieve this by mixing the mode with the vacuum through an unbalanced beam splitter, $\hat{B}(0.1)$. 
We then herald the state on a click in the reflected port and a no-click, c.f. \cref{state-heraldnoclick}, in the transmitted port. 
By doing so, the heralding event occurs with low probability since most of the time the photons in the measured mode
are transmitted and the event discarded. 
However, when we have a successful heralding, there is a high probability that the click in the reflected port was caused by exactly one photon due to the low reflectivity of the beam splitter, and at the same time the no-click in the transmitted port guarantees that no other photon was detected. 
Overall, this simulates a good approximation of the projection on a single photon, provided that the detectors involved are ideal.

\subsection{Results}

We run the random search and reinforcement learning routine over a fixed number of episodes, where each episode terminates when the circuit depth reaches its maximum predefined value, $n_{\rm circuit}$. 
We explored different values of $n_{\rm circuit}$ to verify what violations can be achieved both in circuits with few gates and in circuits with large depth. 
The parameters of the photonic gates are optimized at the end of each episode such that the CHSH score is maximized, with a numerical procedure based on Nelder-Mead optimization. 
Subsequently, we fix the CHSH score and optimize the gates parameters to maximize the total heralding probability of the state. 

In \cref{tab:circuits} and \cref{tab:circuits-agent}, we report a selection of photonic circuits found by random search and reinforcement learning, characterized by their CHSH score and heralding probability. 
For each search strategy, we report only the best circuit across both heralding mechanisms, omitting circuits based on a mechanism that is outperformed by the other.
We focus on circuits with $N=4$ and $N=6$ modes, as our findings suggest that five modes offer no significant advantage over four.

In order to evaluate the various strategies without having to fine tune the RL parameters (e.g.~update frequency, trajectory capacity, clip range, ADAM optimizer, and so on), we perform a random search, where the photonic operations are selected uniformly at random in each step, thereby saving us some time. In this evaluation, strategies 2 and 4 appear suboptimal (c.f. \cref{tab:circuits}), and we thus focus on the other strategies.

\begin{figure}
  \subfloat[Strategy 1, $N=4$ modes, $n_{\rm circuit}=4$ total gates, heralding scheme: single-photon projection.]{\label{fig:learning1}%
  \includegraphics[width=.4\linewidth]{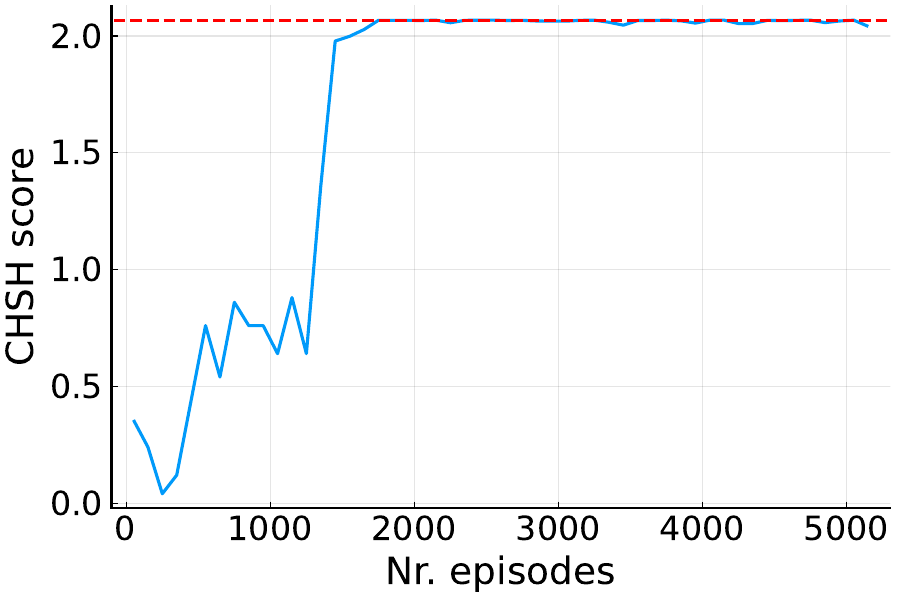}}\qquad
  \subfloat[Strategy 3, $N=6$ modes, $n_{\rm circuit}=12$ total gates, heralding scheme: single-photon projection.]{\label{fig:learning3}%
  \includegraphics[width=.4\linewidth]{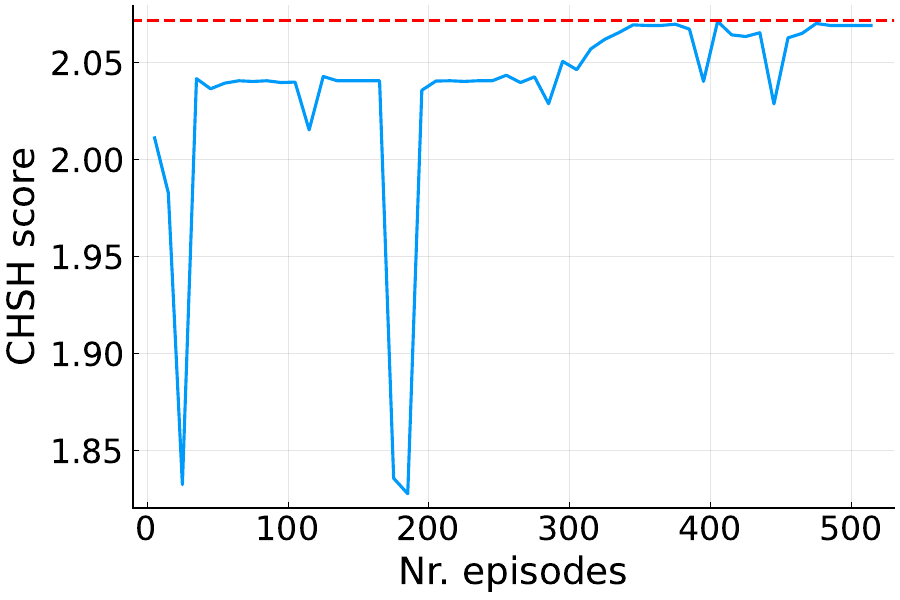}}\qquad
  \subfloat[Strategy 5, $N=6$ modes, $n_{\rm circuit}=20$ total gates, heralding scheme: single-photon projection.]{\label{fig:learning5}%
  \includegraphics[width=.4\linewidth]{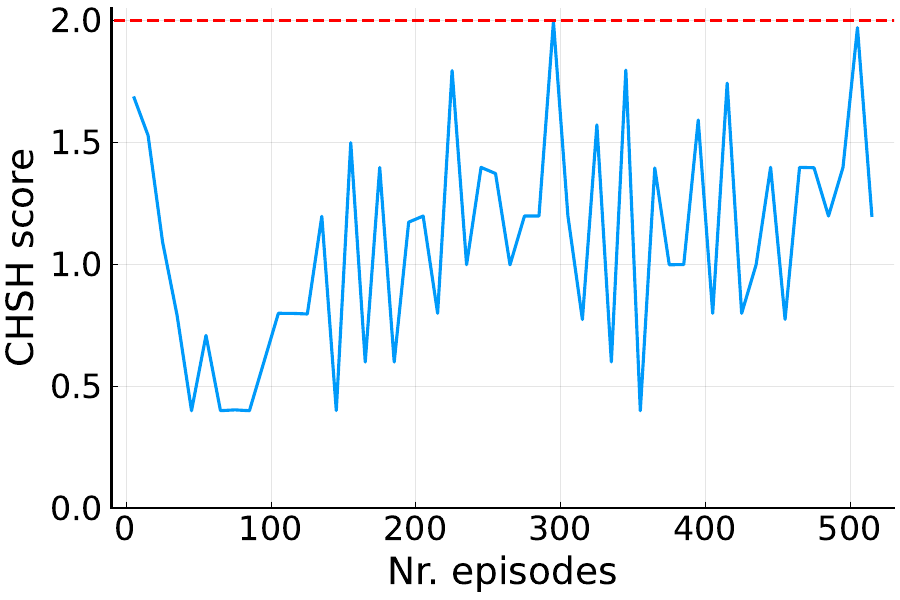}}
\caption{The evolution of the CHSH score with the number of completed episodes, for an agent with PPO policy implementing three different strategies to build the optical circuits (c.f. \cref{app:circuits}). In both (a) and (b), the increase of the CHSH score with the number of episodes shows that the agent learns from previous interactions with the environment to produce circuits with higher CHSH violations. In particular, the agent finds circuits with CHSH score equal to 2.068 (red dashed line) in (a) and 2.072 (red dashed line) in (b). Conversely, the CHSH score remains constantly below 2.0 (red line) in (c), signifying that the agent could not learn a policy leading to CHSH violations. Samples of the circuits found with the three strategies are reported in \cref{tab:circuits-agent}.}
\label{fig:learning}
\end{figure}

The circuits found by the agent in \cref{tab:circuits-agent} are representative of the learning process occurring with the PPO policy trained using the hyperparameters given in \cref{tab:hyperparameters}. In fact, \cref{tab:circuits-agent} reports circuits found by the agent towards the end of the training process, where the CHSH score associated with the circuit becomes larger. The learning process can be visualized with the plots in \cref{fig:learning}, which show the progression of the CHSH score over the training episodes. We observe that, for
Strategy~1 in \cref{fig:learning}(a) and Strategy~3 in \cref{fig:learning}(b), the CHSH score improves over the training and reaches the maximum value in the later episodes. Conversely, it seems that the PPO agent struggles to learn circuit patterns leading to CHSH violations when adopting Strategy~5. This could be explained by the fact that, in Strategy~5, Alice's and Bob's modes are initialized in a product state that would result, if measured, in uncorrelated measurement outcomes
($\braket{A_x B_y}=0$), opposed to the correlations that arise when initializing the modes with a two-mode squeezer ($\braket{A_x B_y}>0$). This means that the starting value of the CHSH score, when the circuit is still empty, with Strategy~5 is closer to zero compared to Strategy~3, where it starts closer to $2$. Thus, it is easier for the agent in Strategy~3 to find circuit patterns leading to CHSH violations than when Strategy~5 is adopted. Still, we observe that the PPO agent in Strategy~5
can find circuits leading to a CHSH score of 2.0, i.e., a policy that generate (trivial) classical correlations between Alice and Bob is learned.

\begin{table}[h]
\centering
\begin{tabular}{ll}
\toprule
\textbf{Parameter} & \textbf{Value} \\
\midrule
Optimizer & Adam ($1\times10^{-4}$) \\
Discount factor $\gamma$ & 0.99 \\
GAE parameter $\lambda$ & 0.95 \\
Clip range & 0.1 \\
Max gradient norm & 0.5 \\
Number of epochs & 25 \\
Number of microbatches & 4 \\
Actor loss weight & 1.0 \\
Critic loss weight & 0.5 \\
Entropy loss weight & 0.001 \\
Update frequency & 64 \\
Trajectory Capacity & 256
\end{tabular}
\caption{Hyperparameters used to train PPO agents as defined in \cite{JuliaRLcommit}.}
\label{tab:hyperparameters}
\end{table}

Finally, training the agent both with the PPO policy and the random search allows us to benchmark the usefulness of a learning agent for circuit generation. By comparing the circuits found in Tables~\ref{tab:circuits} and \ref{tab:circuits-agent}, there is no apparent advantage in using a trained agent to find good candidate circuits, compared to a random search, in this scenario. We remark, however, that this conclusion holds in the context of our RL trainings, which are limited by the computational power of a personal computer and by a shallow neural network with two hidden layers, but we expect a larger neural network to perform better than the one used here. Furthermore, we expect the performance of an unstructured search to be less appealing when trying to construct circuits in larger scenarios, e.g. involving more than two parties.

\begin{table}
\caption{Circuits found by the agent with random search. ``S.'' indicates the strategy adopted, $n_\text{circuit}$ is the number of Gaussian operations, where we indicate in brackets the number of active Gaussian operations (squeezers). The notation is the one adopted in \cref{table:GaussianOperations} and the gates are applied sequentially in the same order as they appear in the list. We omit the $\hat{ }$ symbol in the gates list for ease of notation. The heralding of the modes $3, 4, \dots, N$ is ``cl.'' when the state is heralded on a click in a threshold detector, and ``s.p.'' when attained by simulating a single photon projection; we also report the total heralding probability of the state under ``Herald.''.}
	\label{tab:circuits}
\begin{tabular}{c c c p{0.55\linewidth} p{0.1\linewidth} c }
\toprule
\textbf{S.} & \textbf{$N$} & \textbf{$n_\text{circuit}$} & \textbf{Gates} & \textbf{Herald.} & \textbf{CHSH} \\
\midrule
1 & 4 & 4(2) & $S_2(0.44628)[1,2] \quad B(1.60873)[1,3] \quad S_2(0.03308)[3,4] \newline B(1.60953)[1,3]$  & cl. \newline $5.0 \cdot 10^{-6}$ & 2.068 \\
& 4 & 4(2) & $S_2(0.4455)[1,4] \quad B(3.06797)[1,3] \quad B(1.61219)[2,4] \newline S_2(0.01711)[1,4]$  & cl. \newline $4.9 \cdot 10^{-6}$ & 2.068 \\
\midrule
2 & 4 & 6(1) & $S_2(1.51498)[1,2] \quad B(1.38042)[1,4] \quad B(1.79175)[1,2] \newline B(3.54325)[1,3] \quad B(0.92456)[1,4] \quad B(2.50137)[2,4]$  & s.p. \newline  $4.7 \cdot 10^{-6}$ & 2.060 \\
\midrule
3 & 4 & 5(2) & $S_2(0.00095)[1,2] \quad S_2(0.4501)[3,4] \quad B(1.50217)[2,4] \newline R(0.47822)[4] \quad B(1.63927)[1,3]$ & cl. \newline  $3.2 \cdot 10^{-6}$ & 2.068 \\
 & 4 & 8(2) & $S_2(0.00102)[1,2] \quad S_2(0.44752)[3,4] \quad B(1.62564)[1,2] \newline B(1.49052)[2,3] \quad B(1.31128)[1,2] \quad B(1.63383)[2,4] \newline R(2.10579)[3] \quad B(2.88315)[1,4]$ &  cl. \newline  $3.3 \cdot 10^{-6}$ & 2.068 \\
& 4 & 11(2) & $S_2(0.45013)[1,2] \quad S_2(0.00127)[3,4] \quad B(2.83692)[3,4] \newline B(2.21531)[1,4] \quad B(1.68516)[1,2] \quad B(1.47966)[3,4] \newline B(1.83057)[1,2] \quad B(1.30233)[2,3] \quad B(2.16635)[1,2] \newline R(0.21311)[4] \quad B(1.03997)[2,3]$ &  cl. \newline  $4.1 \cdot 10^{-6}$ & 2.068 \\
& 6 & 9(3) & $S_2(0.4886)[1,2] \quad S_2(0.0103)[3,4] \quad S_2(1.14363)[5,6] \newline B(1.93776)[2,6] \quad R(0.40101)[3] \quad B(1.56028)[5,6] \newline B(2.65334)[4,5] \quad B(2.1377)[3,6] \quad B(1.79887)[1,3]$ & s.p. \newline  $1.8 \cdot 10^{-11}$ & 2.073 \\
\midrule
4 & 4 & 9(2) & $S_1(1.57601)[1] \quad S_1(0.30398)[2] \quad R(1.57557)[2] \newline B(1.38422)[1,3]  \quad B(1.44189)[1,4] \quad B(0.20825)[2,3] \newline B(1.75776)[3,4] \quad B(1.54447)[1,4]  B(2.75608)[1,2]$ & cl. \newline $1.2 \cdot 10^{-8}$ & 2.063 \\
\midrule
5 & 4 & 17(4) & $S_1(1.37645)[1] \quad S_1(0.00168)[2] \quad S_1(0.00599)[3] \newline S_1(0.22444)[4]  \quad  R(0.00749)[3] \quad B(0.58858)[1,2] \newline R(0.54082)[4] \quad B(3.14062)[2,3]  \quad B(2.69734)[1,3] \newline B(2.30103)[1,2] \quad B(1.56934)[3,4] \quad R(1.02956)[3]  \newline B(0.94574)[1,4] \quad B(4.25384)[1,3] \quad B(1.31202)[2,3] \newline B(1.74175)[1,2] \quad B(2.5986)[1,4] $  & cl. \newline $1.2 \cdot 10^{-5}$ & 2.066 \\
& 4 & 19(4) & $S_1(0.70577)[1] \quad S_1(0.00066)[2] \quad S_1(0.4052)[3] \newline S_1(0.06303)[4]  \quad B(1.29941)[2,3] \quad B(1.4099)[1,3] \newline B(1.02284)[2,3] \quad R(0.91296)[1]  R(0.95323)[3] \quad \newline \quad R(1.44978)[3] \quad B(1.97387)[1,3]  B(2.60931)[2,4] \newline R(0.43968)[4] \quad B(1.37219)[1,2] \quad R(0.62318)[3]  \newline \quad B(2.97856)[1,3] \quad B(2.17853)[1,2] $ & s.p. \newline  $2.1 \cdot 10^{-7}$ & 2.068\\
& 4 & 11(4) & $S_1(1.99879)[1] \quad S_1(0.27027)[2] \quad S_1(0.02467)[3] \newline S_1(0.14767)[4] \quad B(1.5709)[1,2] \quad R(1.57177)[1] \newline B(1.19975)[1,3] \quad B(2.85582)[2,3] \quad B(1.94903)[3,4] \newline B(1.62179)[1,4] \quad B(1.71619)[1,2]$ & s.p. \newline $ 2.3 \cdot 10^{-7}$ & 2.068 \\
\multicolumn{6}{r}{(Continued on next page ...)}
\end{tabular}
\end{table}

\begin{table}
    \centering
    \begin{tabular}{c c c p{0.55\linewidth} p{0.1\linewidth} c}
    \multicolumn{6}{c}{Table 4 -- Continued from previous page} \\
    \toprule
    \textbf{S.} & \textbf{$N$} & \textbf{$n_\text{circuit}$} & \textbf{Gates} & \textbf{Herald.} & \textbf{CHSH} \\
    \midrule
    & 6 & 27(6) & $S_1(0.13442)[1] \quad S_1(0.01894)[2] \quad S_1(0.23412)[3] \newline S_1(0.75517)[4]  \quad S_1(1.91139)[5] \quad S_1(0.06771)[6] \newline B(0.77041)[1,4] \quad B(1.44169)[1,2] \quad B(0.84147)[1,4] \newline R(0.33784)[4] \quad B(1.05774)[5,6] \quad B(1.41175)[2,5] \newline B(1.0452)[2,4] \quad B(1.83876)[2,6] \quad B(1.59497)[1,3] \newline R(1.52075)[4] \quad B(0.77169)[1,2] \quad R(0.69784)[2] \newline B(1.34197)[1,3] \quad B(1.67829)[1,4] \quad B(2.63474)[2,3] \newline
R(1.42985)[2] \quad B(1.90975)[3,5] \quad B(2.04856)[4,5] \newline R(2.17427)[5] \quad B(2.3683)[2,3] \quad B(0.98299)[1,4]$ & s.p. \newline  $ 3.5 \cdot 10^{-12}$ & 2.070\\
& 6 & 42(6) & $S_1(1.33522)[1] \quad S_1(0.50981)[2] \quad S_1(1.35176)[3] \newline S_1(1.78416)[4] \quad S_1(1.56906)[5] \quad S_1(0.14952)[6] \newline B(1.45766)[2,3] \quad B(1.12)[2,5]  \quad B(1.13792)[3,6] \newline R(1.21553)[6] \quad B(1.61669)[4,6] \quad R(0.52559)[2] \newline B(1.40038)[3,6] \quad B(1.89262)[2,6] \quad B(1.80117)[4,5] \newline R(0.58755)[6] \quad B(2.19939)[1,5] \quad B(1.41454)[4,6] \newline B(2.14052)[3,5] \quad R(0.49489)[2] \quad B(1.97156)[5,6]
\newline R(0.42675)[1] \quad B(2.04066)[2,5] \quad B(1.88537)[3,6] \newline B(2.04211)[2,5] \quad B(1.75719)[4,5] \quad B(1.39104)[1,2] \newline B(1.84995)[2,6] \quad B(1.24971)[5,6] \quad R(0.5619)[6] \newline B(1.20624)[2,3] \quad R(0.17554)[5] \quad B(1.29781)[2,3] \newline R(0.98721)[2] \quad R(0.30125)[1] \quad B(1.76338)[1,2] \newline B(1.66911)[3,6] \quad B(1.26734)[1,5] \quad R(0.35314)[4] \newline B(1.47)[1,4] \quad B(1.25175)[1,3] \quad R(1.80732)[4]$ & s.p. \newline  $ 1.7 \cdot 10^{-11}$ & 2.076\\
\bottomrule
    \end{tabular}
\end{table}

\begin{table}
\caption{Circuits found using reinforcement learning with the PPO policy. The same notation of \cref{tab:circuits} applies.}
\label{tab:circuits-agent}
	\begin{tabular}{c c c p{0.55\linewidth} p{0.1\linewidth} c }
	\toprule
	\textbf{S.} & \textbf{$N$} & \textbf{$n_\text{circuit}$} & \textbf{Gates} & \textbf{Herald.} & \textbf{CHSH} \\
	\midrule
	1 & 4 & 4(3) & $S_2(0.58994)[1,2] \quad S_2(1.13938)[3,4] \quad B(2.83528)[2,4] \newline S_2(1.15427)[3,4]$  &  s.p. \newline $3.0 \cdot 10^{-6}$  & 2.068 \\
	1 & 4 & 4(2) & $S_2(0.00462)[3,4] \quad B(1.62748)[2,4] \quad S_2(0.44529)[1,2] \newline B(3.01595)[2,4]$  & s.p. \newline $2.1 \cdot 10^{-11}$ & 2.069 \\
	\midrule
	3 & 4 & 4(2) & $S_2(0.00096)[1,2] \quad S_2(0.44993)[3,4] \quad B(1.63856)[2,4] \newline B(1.50272)[1,3]$ & cl. \newline $3.0 \cdot 10^{-6}$ & 2.068 \\
	3 & 6 & 8(3) & $S_2(1.00266)[1,2] \quad S_2(0.77432)[3,4] \quad S_2(0.05986)[5,6] \newline B(3.1033)[4,6] \quad B(2.90802)[2,3] \quad B(1.13535)[4,6] \newline B(2.84339)[2,5] \quad B(0.38529)[2,3] $ & s.p. \newline $1.1 \cdot 10^{-11}$  & 2.072 \\
	\midrule
5 & 4 & 10(4) & $S_1(0.97488)[1] \quad S_1(0.03222)[2] \quad S_1(1.16719)[3]  \newline \quad  R(1.56407)[1] \quad B(1.39216)[3,4] \quad B(2.53958)[1,4] \newline B(1.28743)[1,2] \quad B(1.09283)[1,4] \quad B(2.36714)[3,4]$ & s.p. \newline $2.3 \cdot 10^{-6}$  & 2.030 \\
5 & 6 & 20(6) & $S_1(1.8)[1] \quad S_1(1.8)[2] \quad S_1(1.0)[3] \newline S_1(1.0)[4]  \quad  S_1(1.0)[5] \quad  S_1(1.0)[6] \newline R(1.57318)[1]  \quad B(2.35548)[1,2]  \quad R(0)[1] \newline R(0)[1] \quad R(0)[1] \quad R(0)[1] \newline R(0)[1] \quad R(0)[1] \quad R(0)[1] \newline R(0)[1]\quad R(0)[1]\quad R(0)[1] $ & s.p. \newline  $1.2 \cdot 10^{-19}$  & $1.996$ \\
\bottomrule
\end{tabular}
\end{table}

\clearpage

\section{Photonic circuit for Bell test with homodyne measurement}
\label{app:robustness}

In this section, we provide details on our analysis of the simple photonic depicted in \cref{fig:result1}.

\subsection{Robustness to photon loss}
\label{app:photonloss}

We here study the impact of photon loss on the CHSH score. We first assume a \textit{direct transmission} configuration in which Alice prepares the heralded state before sending a subsytem to Bob, i.e. state preparation including the heralded modes are kept within Alice's laboratory.
This configuration is arguably in-line with what is expected for setup used in device-independent protocols or for a scalable quantum Internet where quantum nodes are connected in a decentralized way. Moreover, the simplicity of this configuration, requiring only two labs, is privileged for the realization of a first proof-of-concept experiments.

We assume that photon loss originates solely from attenuation in the optical fiber connecting Alice to Bob. We consider a standard attenuation of $\gamma =  0.2$dB/km.
The total optical transmittance, i.e. the ratio of transmitted photons, over a distance $L$ is given by
\begin{equation}
    T = 10^{-\gamma L/10}
\end{equation}
This transmittance can be simply modeled with a beam splitter of transmittance $T$, coupling the bosonic mode of Bob to an empty bosonic mode which can be traced out.
In term of beam-splitter angle parameter $\theta$, as per \cref{table:GaussianOperations}, this is $\theta=\arccos(\sqrt{T}).$

We optimize the parameters of the different gates of the circuits for every increasing distance, until a CHSH score lower than $2+10^{-8}$ is encountered. 
Starting at distance $L=0$km, we set the parameters to ones given in the main text which lead to a score of $2.068$. Parameters are then optimized using the \textsc{NelderMead} algorithm, for distance increasing by step of $0.5$km until $8.0$km and by step of $0.1$ afterwards, ensuring a finer analysis for the high photon loss regime. 
Note that the parameters are constrained within experimentally realistic values : squeezing parameters are kept in the $[0,1.15]$ range, i.e. below $10$dB of squeezing.
In \cref{tab:parameters} we give the optimized parameters of the different optical elements as well as the resulting CHSH score for distance between $0$km and $8.1$km.

\begin{table}
\centering
\caption{Parameters of the photonic gates optimizing the CHSH score for different distances given in kilometers. $r_1$ and $r_2$ are the squeezing parameters of the two-mode squeezers on mode $1,2$ and $3,4$ respectively. $\theta_1$ and $\theta_2$ denote the angle of the beam-splitter between mode $1,3$ and $2,4$ respectively. We note that the squeezing parameter $r_2$ is saturating the upper constraint of $1.15$ indicating that higher violation and a better robustness to photon loss are to be expected if one would allow for a greater squeezing range. }
\label{tab:parameters}
	\begin{tabular}{c c c c c c }
    \toprule
    Distance (km) & CHSH score & $r_1$ & $r_2$ & $\theta_1$ & $\theta_2$ \\
    \midrule
        0.0 & 2.06817 & 4.48373e-5 & 0.441672 & 1.56714 & 1.62989 \\
        0.5 & 2.06041 & 2.1759e-5 & 0.461487 & 1.56899 & 1.62893 \\
        1.0 & 2.05303 & 1.47679e-5 & 0.48497 & 1.56959 & 1.63078 \\
        1.5 & 2.04608 & 1.46126e-5 & 0.508651 & 1.56958 & 1.63013 \\
        2.0 & 2.03972 & 2.13881e-5 & 0.534306 & 1.56783 & 1.60718 \\
        2.5 & 2.03365 & 1.54105e-5 & 0.563286 & 1.56872 & 1.60932 \\
        3.0 & 2.0281 & 1.40263e-5 & 0.591939 & 1.5477 & 1.57405 \\
        3.5 & 2.02299 & 1.23683e-5 & 0.624187 & 1.54947 & 1.57404 \\
        4.0 & 2.0184 & 1.05311e-5 & 0.660672 & 1.55651 & 1.57517 \\
        4.5 & 2.01432 & 9.35385e-6 & 0.7029 & 1.56595 & 1.58322 \\
        5.0 & 2.01075 & 5.47606e-6 & 0.74912 & 1.55381 & 1.5731 \\
        5.5 & 2.00774 & 6.03821e-6 & 0.799573 & 1.56625 & 1.58154 \\
        6.0 & 2.00526 & 5.05152e-6 & 0.85906 & 1.56403 & 1.57805 \\
        6.5 & 2.00331 & 4.57833e-6 & 0.929927 & 1.56254 & 1.57782 \\
        7.0 & 2.00187 & 2.4149e-6 & 1.00901 & 1.56177 & 1.57586 \\
        7.5 & 2.0009 & 6.45471e-7 & 1.1212 & 1.56417 & 1.5759 \\
        7.6 & 2.00075 & 2.45411e-7 & 1.13324 & 1.56941 & 1.58318 \\
        7.7 & 2.00062 & 3.04154e-6 & 1.14998 & 1.58628 & 1.58565 \\
        7.8 & 2.00049 & 3.41435e-6 & 1.15 & 1.57531 & 1.5983 \\
        7.9 & 2.00035 & 2.77572e-6 & 1.14998 & 1.57346 & 1.609 \\
        8.0 & 2.00022 & 7.64151e-7 & 1.14991 & 1.57015 & 1.60449 \\
        8.1 & 2.00006 & 2.15693e-7 & 1.15 & 1.57865 & 1.6037
    \end{tabular}
\end{table}

\medskip

For completeness, we further study a \textit{central station} configuration, making use of a third party location, in which the state preparation occurs before being send to Alice and Bob. This configuration may lead to an increased robustness to photon loss caused by distance at the cost of a more complex experimental realization. This is the case for the qubit case where the total efficiency $\eta = \eta_A \eta_B$ required for a CHSH violation is lower for in the symmetric case $\eta_A=\eta_B$ than in the asymmetric case where $\eta_A = 1$~\cite{Brunner2007,Cabello2007}.

We focus on the symmetric case where Alice and Bob are both at a distance $L$ of a central station where state preparation is performed. We use the same model for photon loss than in the \textit{direct transmission} configuration, i.e. each party sees their mode coupled to an empty mode via a beam-splitter with transmittance accounting for photon loss of a lossy fiber at distance $L$. We optimize the circuit parameters for every step of $0.2$km of total distance, with the constraints previously mentioned.

\begin{figure}
    \centering
    \includegraphics[width=0.5\linewidth]{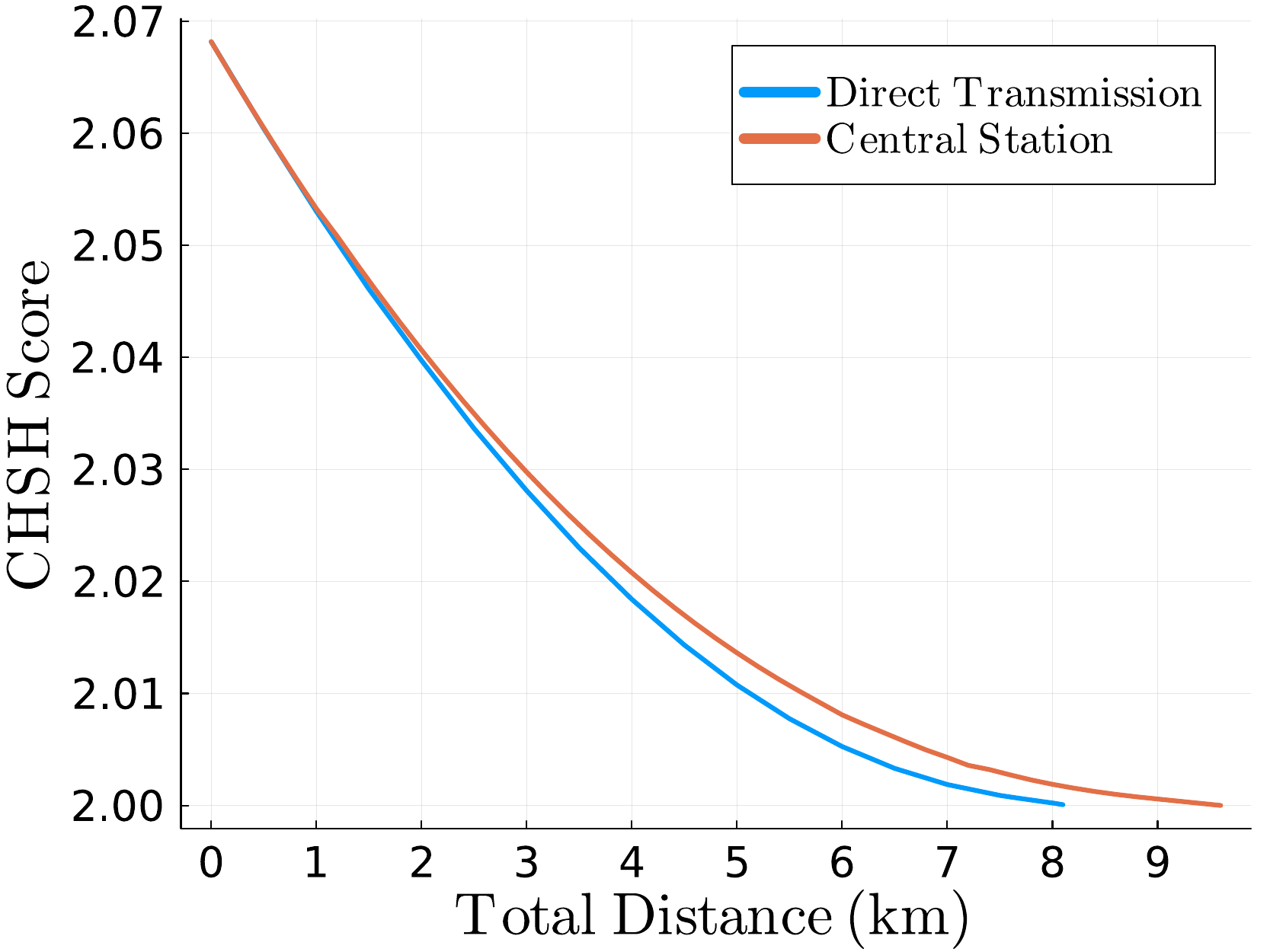}
    \caption{CHSH score with respect to the distance between Alice and Bob. Direct transmission configuration has Alice prepare and send the state to Bob. Central station configuration has a third party preparing the state and sending it to Alice and Bob -- here we assume the case where Alice and Bob are at equidistance from the central station, i.e. half the total distance.}
    \label{fig:resultDistance}
\end{figure}

In \cref{fig:resultDistance}, we plot the CHSH score with the total distance between Alice and Bob in both configuration cases. The central station setup provides more robustness to photon loss, especially at higher distance. Specifically, this configuration allow to separate party further, up to $9.8$km. However, for smaller distance, there is no notable difference between these two configurations.

\subsection{Robustness to inefficient threshold photon-detectors}
\label{app:ineffthreshold}

\begin{figure}
    \centering
    \includegraphics[width=0.5\linewidth]{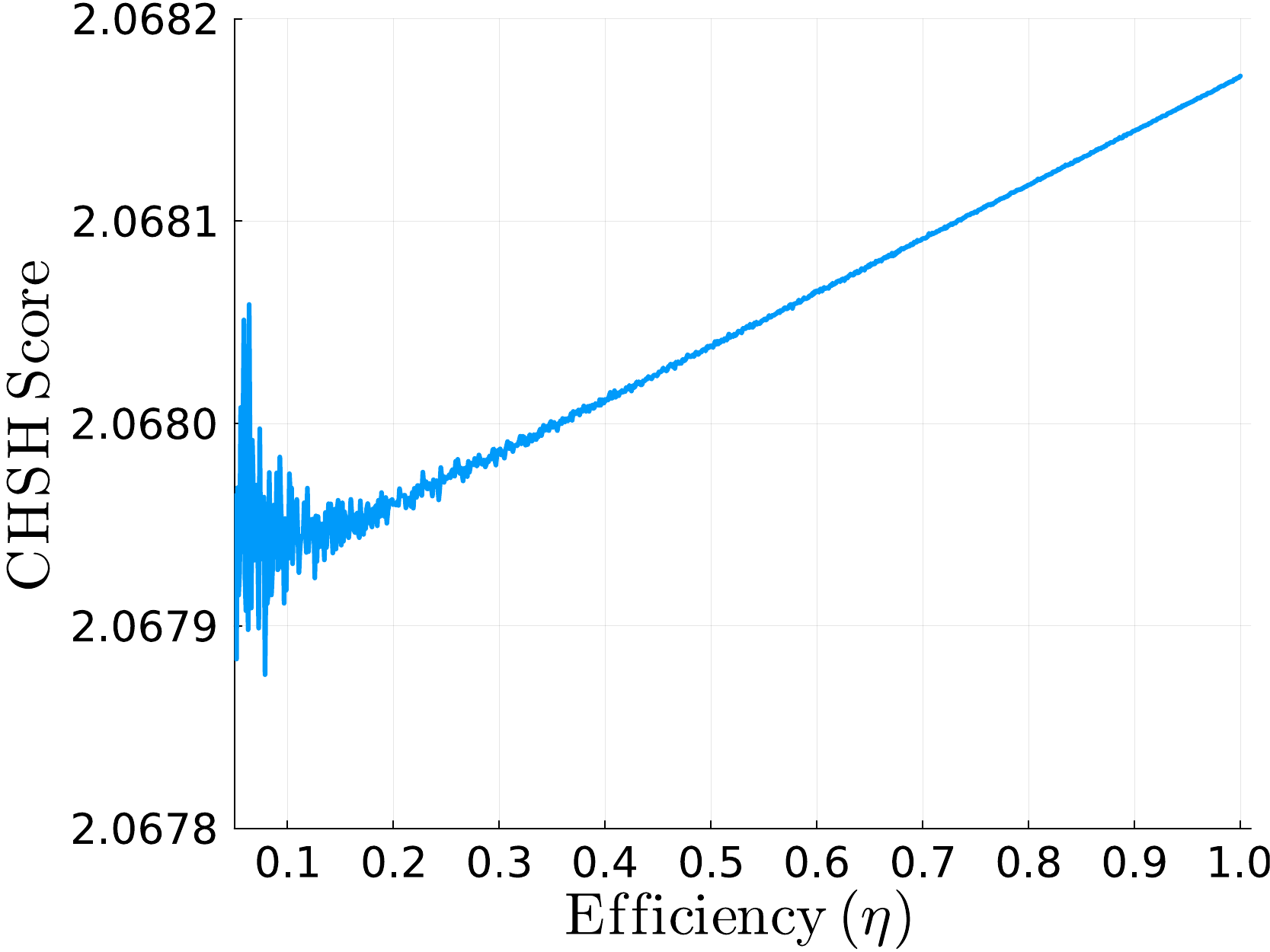}
    \caption{CHSH score with respect to the efficiency of the threshold detectors $\eta$.}
    \label{fig:chshThreshold}
\end{figure}

We analyze the impact of inefficient threshold detectors used in the heralding mechanism, on the CHSH score and on the heralding probability.
We use formulas \cref{herald-prob-noclick} and \cref{state-heraldclick} to compute the probability of a click event of a threshold photon-detector with efficiency $\eta$ and the resulting heralded state.
We fix the parameters of the two squeezers and two beam-splitter to the ones provided in the main text. We consider the two heralded detectors to have the same efficiency $\eta$.
The CHSH score evolution with efficiency $\eta$ is given in \cref{fig:chshThreshold}. We notice that the score is almost constant with $\eta$. This behavior is expected as the state resulting from the heralding mechanism is marginally affected by the efficiency of the heralding detectors.
Note that we set a limit $\eta>0.05$ as lower efficiency introduces numerical error due to the ratio appearing in \cref{state-heraldclick}, exploding for low detection probability.

\end{document}